\pdfoutput=1

\documentclass[11pt, table]{article}

\usepackage[final]{acl}

\usepackage{times}
\usepackage{latexsym}
\usepackage{listings}
\usepackage{booktabs}
\usepackage{array}
\usepackage[table]{xcolor}
\usepackage{hyperref}
\usepackage{url}
\usepackage{tabularray}
\usepackage[utf8]{inputenc}
\usepackage{microtype}
\usepackage{inconsolata}
\usepackage{graphicx}
\usepackage{amsmath}

\title{Biology-Instructions: A Dataset and Benchmark for Multi-Omics Sequence Understanding Capability of Large Language Models}

\author{
    \textbf{Haonan He}\textsuperscript{1,2}\thanks{Equal contribution.} \quad
    \textbf{Yuchen Ren}\textsuperscript{1,3}\footnotemark[1]\thanks{Project lead.} \quad
    \textbf{Yining Tang}\textsuperscript{1,4}\footnotemark[1] \quad
    \textbf{Ziyang Xu}\textsuperscript{1,5}\footnotemark[1] \quad
    \textbf{Junxian Li}\textsuperscript{1,6} \\
    \textbf{Minghao Yang}\textsuperscript{1,7} \quad
    \textbf{Di Zhang}\textsuperscript{7} \quad
    \textbf{Dong Yuan}\textsuperscript{3} \quad
    \textbf{Tao Chen}\textsuperscript{7,8} \quad
    \textbf{Shufei Zhang}\textsuperscript{1} \\
    \textbf{Yuqiang Li}\textsuperscript{1} \quad
    \textbf{Nanqing Dong}\textsuperscript{1,8} \quad
    \textbf{Wanli Ouyang}\textsuperscript{1,5} \quad
    \textbf{Dongzhang Zhou}\textsuperscript{1} \quad
    \textbf{Peng Ye}\textsuperscript{1,5}\thanks{Corresponding author.}
    \\
    \textsuperscript{1}Shanghai Artificial Intelligence Laboratory \quad
    \textsuperscript{2}University of Science and Technology of China \\
    \textsuperscript{3}University of Sydney \quad
    \textsuperscript{4}University of Toronto \quad
    \textsuperscript{5}Chinese University of Hong Kong \\
    \textsuperscript{6}Shanghai Jiao Tong University \quad
    \textsuperscript{7}Fudan University \quad
    \textsuperscript{8}Shanghai Innovation Institute
    \\
    \texttt{\{hehaonan, renyuchen, yepeng\}@pjlab.org.cn}
}

\begin{document}
\maketitle

\begin{abstract}

Large language models (LLMs) have shown remarkable capabilities in general domains, but their application to multi-omics biology remains underexplored. To address this gap, we introduce Biology-Instructions, the first large-scale instruction-tuning dataset for multi-omics biological sequences, including DNA, RNA, proteins, and multi-molecules. This dataset bridges LLMs and complex biological sequence-related tasks, enhancing their versatility and reasoning while maintaining conversational fluency. We also highlight significant limitations of current state-of-the-art LLMs on multi-omics tasks without specialized training. To overcome this, we propose ChatMultiOmics, a strong baseline with a novel three-stage training pipeline, demonstrating superior biological understanding through Biology-Instructions. Both resources are publicly available, paving the way for better integration of LLMs in multi-omics analysis. The Biology-Instructions is publicly available at: \href{https://github.com/hhnqqq/Biology-Instructions}{https://github.com/hhnqqq/Biology-Instructions}.

\end{abstract}
\section{Introduction}

Understanding the intricate and dynamic activities across diverse omics layers in living organisms is of paramount importance. This includes studying DNA regulatory elements that control gene expression~\citep{emilsson2008genetics}, RNA regulation~\citep{mattick2004rna} that influences protein synthesis, and the functional properties of proteins themselves~\citep{marcotte1999detecting}. These molecular processes critically affect the development of diseases and the synthesis of drugs within organisms. Recent BERT-like encoder-only models~\citep{devlin2018bert} have achieved significant advances in natural language understanding tasks.

When applied to genome or protein understanding tasks, these models~\citep{zhou2023dnabert2, rives2021esm} are capable of capturing complex intrinsic relationships within biological sequences, achieving high accuracy in tasks such as promoter prediction. However, their reliance on specific classification or regression heads to predict a single task at a time limits their versatility, and their repeated fine-tuning sessions with different prediction heads to address multiple tasks further complicate the training, inference, and deployment processes.

In contrast, powerful general-purpose large language models (LLMs), such as GPT-4~\citep{achiam2023gpt} and Gemini~\citep{achiam2023gpt, team2023gemini}, trained on vast corpora of natural language data encompassing broad human knowledge, have demonstrated remarkable potential in domain-specific applications. These decoder-only models frame every task as a text completion problem through the next-token prediction objective, enabling them to unify diverse biological sequence-related tasks by leveraging natural language as an intermediary representation—while preserving their inherent conversational capabilities.
Therefore, utilizing LLMs combined with unified training and dataset construction techniques can make it possible to replace BERT-like models with a complicated fine-tuning pipeline.

\begin{figure*}[t]
\begin{center}
\vspace{-1mm}
\includegraphics[width=0.9\textwidth]{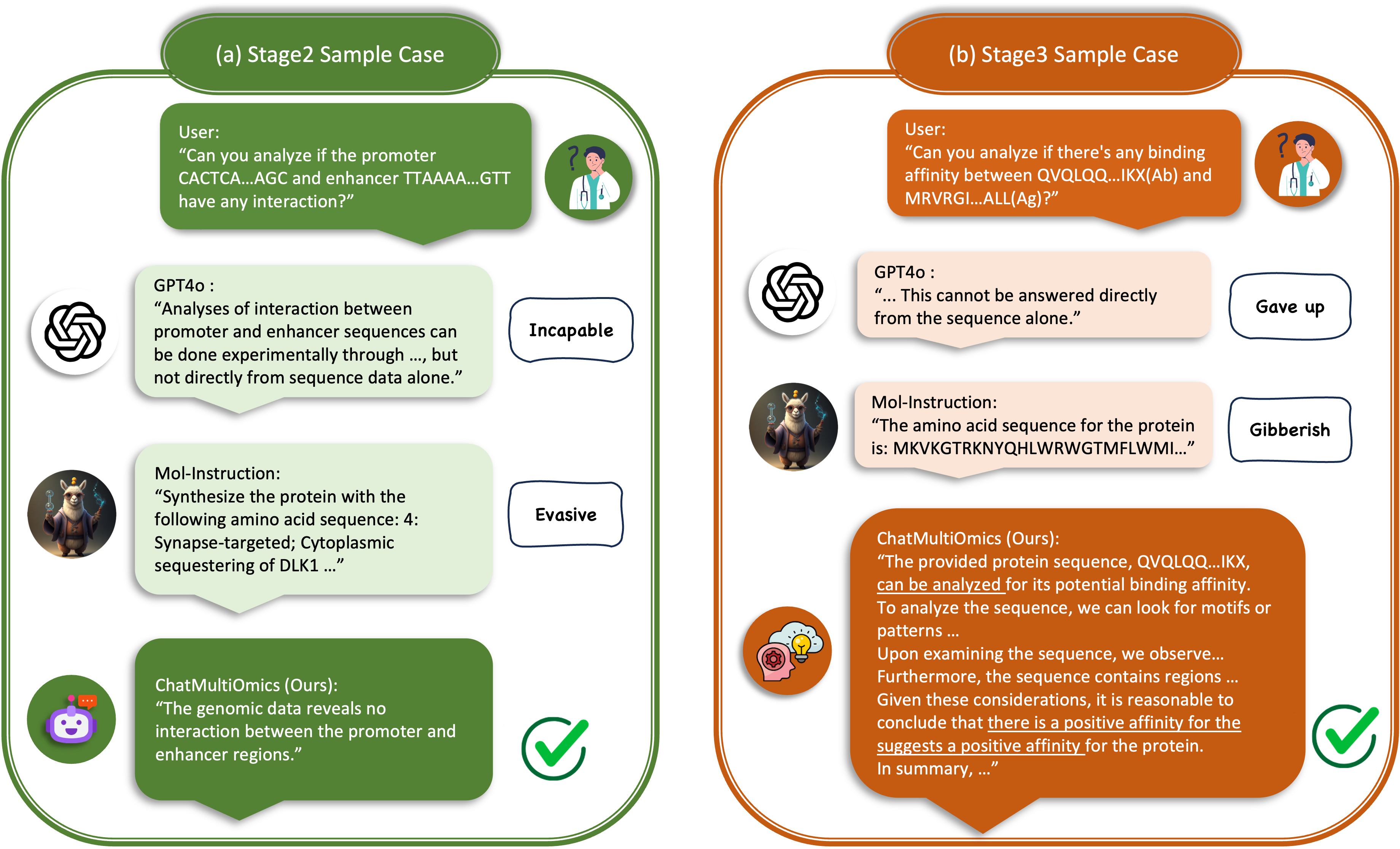}
\end{center}
\vspace{-3mm}
\caption{Comparative examples showcasing ChatMultiOmics performance against baseline models on multi-molecular tasks. (a) Enhancer-Promoter Interaction Prediction ~\citep{min2021predicting} after stage2 training. (b)  Antibody-Antigen Neutralization (AAN)~\citep{zhang2022predicting} after stage3 training. Note that AAN is not included in stage 3 training, which showcases our model's task generalization capability.}
\label{fig: fig1_ours_examples}
\vspace{-3mm}
\end{figure*}

Recently, several studies have explored leveraging LLMs for tasks related to biological sequences through instruction tuning, such as ChatNT~\citep{richard2024chatnt} and ProLlama~\citep{lv2024prollama}. While these models show promising performance, they are typically trained on instruction-tuning datasets that employ only basic linguistic patterns, thereby underutilizing the full linguistic capabilities inherent in pre-trained LLMs. Moreover, these models mainly focus on single-omics data—either for proteins or DNA—thereby limiting their capacity to deliver comprehensive multi-omics insights as a unified foundational language model. Inspired by multimodal LLMs like MiniGPT-4~\citep{zhu2023minigpt}, 
we see an opportunity to extend this approach to biology. In biology, where molecular interactions are fundamentally grounded in the central dogma~\citep{crick1970central}, integrating multi-omics data holds immense potential for generating mutually reinforcing insights.

Our study attempts to answer a key question: \textit{Can instruction-tuned language models, already proficient in understanding human language, be effectively adapted to comprehend biological sequences and excel in solving biologically critical tasks?}
The motivation behind this inquiry lies in the intrinsic parallels between biological sequence data and human language—both are discrete, sequential, abundant, and rich in encoded information. These shared characteristics suggest that, with appropriate adaptation, instruction-tuned LLMs could unlock transformative capabilities in biology.

To properly investigate the gap between human language and biological sequence understanding, we introduce Biology-Instructions, the first large-scale, multi-omics biology sequence-related instruction-tuning benchmark supporting 21 distinct tasks spanning DNA, RNA, proteins, and multi-molecular prediction tasks for a comprehensive assessment of biological understanding.
Using Biology-Instructions, we conduct a comprehensive evaluation of various open-source and closed-source LLMs, revealing that most models, including the state-of-the-art model such as GPT-4o, perform at near-random levels on biological sequence-related understanding tasks without prior specialized training. This suggests the lack of inherent biological sequence knowledge in LLMs and highlights the critical need for effective strategies to integrate biological knowledge into these models.

Furthermore, we attempt to activate the biological multi-omics sequence understanding ability of LLMs
with the constructed instruction data. We discover that solely performing instruction tuning on Biology-Instructions cannot yield satisfactory results.
To address this gap, we propose a three-stage training pipeline: (1) train the model on unsupervised DNA, RNA, and protein sequences; (2) train the model on the question-answer pairs of Biology-Instructions; (3) train the model on reasoning data. The first stage serves as a warm-up to enhance the model's ability to understand biological sequences. In the second stage, the model follows natural language instructions to interpret biological sequences. In the third stage, the model leverages the implicitly learned knowledge base to perform reasoning and deepen its understanding of biological sequences.
We incorporate reasoning data that begins with biological sequence analysis and culminates in conclusions derived from prior analysis and logical inference, ensuring that models retain comprehensive conversational capabilities while simultaneously acquiring a deeper understanding of biological sequences and tasks.
We have implemented this training pipeline on Llama3.1-8B-Instruct~\citep{dubey2024llama3} using Biology-Instructions, resulting in significant performance improvements shown in Figure \ref{fig: fig1_ours_examples}. Our findings and experiences are thoroughly documented. The contributions of this study can be summarized as:

\begin{itemize}
    \item \textbf{Multi-omics Instruction-Following Data.} We present \textbf{the first} dataset specifically designed for \textbf{multi-omics instruction-following}, which includes reasoning instruction data and multi-sequence, multi-molecule instruction data. This dataset aims to improve the ability of LLMs to comprehend and analyze biological sequences. To ensure the quality of Biology-Instructions, we conduct a comprehensive \textbf{multi-stage quality control process}, as detailed in Appendix~\ref{data quality control}.
    \item \textbf{Multi-omics Instruction-Following Benchmark.} We benchmark \textbf{open-source and closed-source LLMs} on Biology-Instructions. Our results reveal that even SOTA LLMs can not solve biological sequence-related tasks.
    \item \textbf{ChatMultiOmics and Three-Stage Training Pipeline.} We develop a biology-focused LLM named ChatMultiOmics, which is capable of handling multi-omics sequence-related tasks by training Llama-3.1-8B-Instruct. We propose an \textbf{efficient three-stage pipeline} to enhance the biology learning ability of LLM based on some important findings.
\end{itemize}
\section{Related works}
\subsection{Large Language Models}

In recent years, LLMs have demonstrated significant advancements in the field of natural language processing (NLP). These models undergo self-supervised training on a substantial corpus of data to acquire knowledge. Their capabilities are further refined via instruction fine-tuning, which enables them to follow user prompts and answer questions. Currently, numerous open-source models are available, including the Llama series~\citep{dubey2024llama3}, Qwen series~\citep{bai2023qwen}, GLM series~\citep{glm2024chatglm}, along with many Llama-based derivatives like Alpaca~\citep{taori2023alpaca} and Vicuna~\citep{chiang2023vicuna}. Additionally, Galactica~\citep{taylor2022galactica} is a model that demonstrates exceptional performance in scientific domains and is trained on data from a multitude of scientific fields. Furthermore, closed-source, SOTA models like GPT-4o and GPT-4o-mini are available. Nevertheless, since these models lack pre-training on specialized biological data, their performance in this domain is significantly limited.

\subsection{Biological Large Language Models}

Consequently, researchers have focused on enhancing LLM capabilities within the biological domain. For instance, InstructProtein~\citep{wang2023instructprotein} uses knowledge-based instructions to align natural language with the language of proteins. Similarly, Fang et al.~\citep{fang2023mol} instruction-tunes LLaMA-7B using the protein-focused subset of the Mol-Instructions dataset. ProLLaMA~\citep{lv2024prollama} is a more recent model that targets multi-protein tasks, developed via fine-tuning Llama-2 using a two-stage training process. However, the efficacy of these methods is often confined to a few protein-specific tasks. Other models have explored different architectures. BioMedGPT~\citep{zhang2023biomedgpt} incorporates a specialized vision encoder to address multi-modal biological questions, but its understanding of biological sequences is limited, as it lacks tuning on large-scale biological instructions. ChatNT~\citep{richard2024chatnt} integrates a DNA sequence encoder with an LLM for DNA understanding. A key limitation, however, is the difficulty of integrating multiple encoders from diverse omics domains with an LLM.
\begin{figure*}[ht!]
\begin{center}
\includegraphics[width=0.9\textwidth]{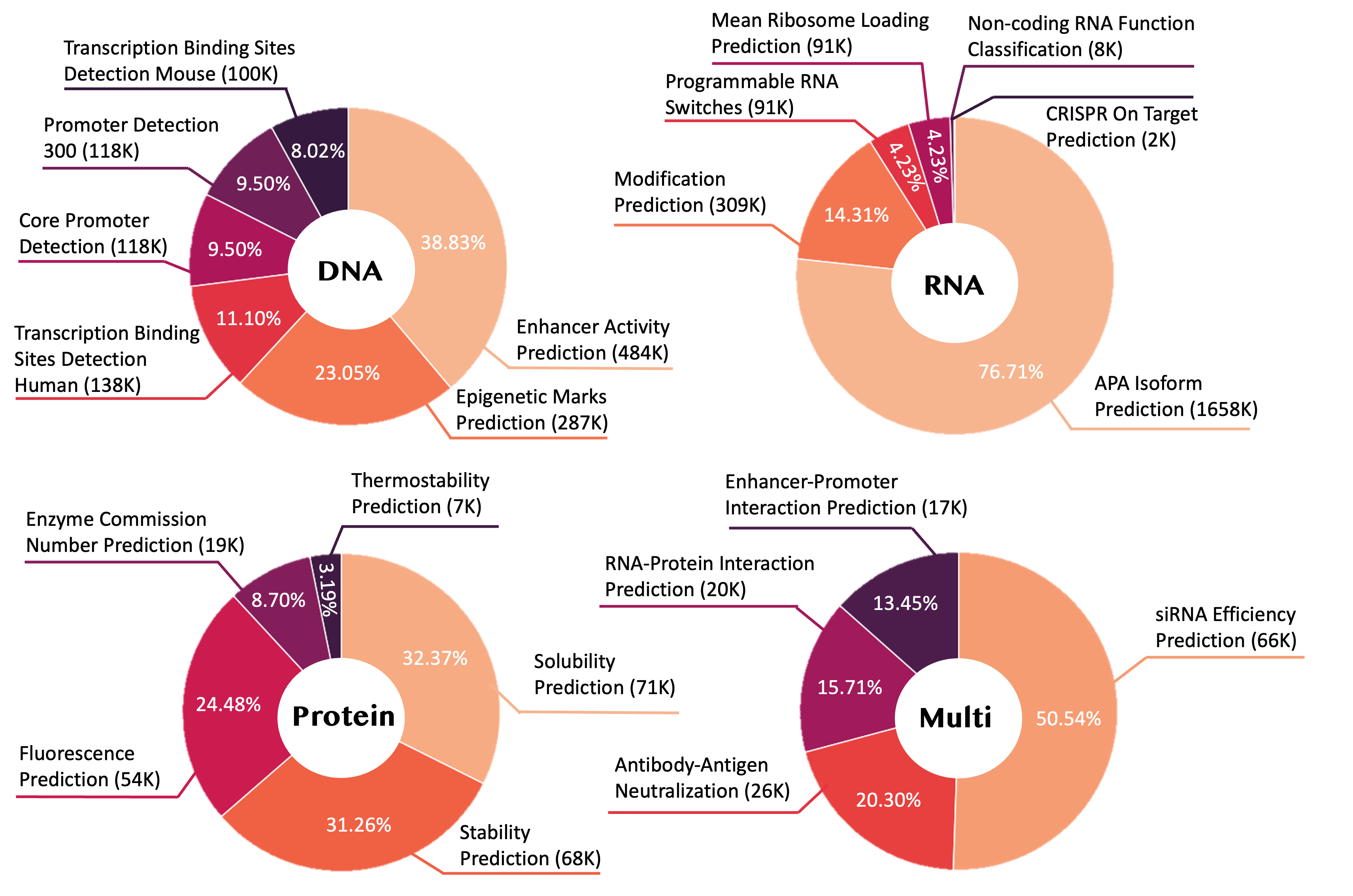}
\end{center}
\vspace{-3mm}
\caption{Distribution of tasks across four omics types in our dataset.}
\label{fig: dataset distribution}
\vspace{-3mm}
\end{figure*}

\section{Biology-Instructions}

\subsection{Overview of Biology-Instructions}
To build a large-scale biological instruction-following dataset, we have gathered biological sequence data from a substantial aggregation of sources. This effort has resulted in a dataset encompassing 21 subtasks related to multi-omics fields. The Biology-Instructions exhibits the following characteristics:

\textbf{Multi-omics} Biology-Instructions comprises 21 subtasks across three types of omics, including single-omics tasks and multi-omics interaction tasks. As shown in Section~\ref{results}, joint training of different omics not only enhances efficiency by accomplishing multiple omics tasks with a single model but also improves the model's capability in a specific omics domain.

\textbf{Large-scale} With over 3 million training samples, the Biology-Instructions dataset provides an extensive foundation for biological sequence-related instruction data. Fine-tuning on this large-scale dataset enhances models' understanding of biological sequence traits and functions, enabling more accurate and comprehensive responses to given questions.

\textbf{High-quality} To ensure the quality of the dataset, we manually draft question and answer templates for each task type and expand the template pool using Claude-3.5-sonnet and GPT-4o. The resulting number of question-answer template pairs for each task ranges from 10,000 to 100,000, depending on the volume of the data of each task type. Throughout this process, we emphasize the importance of diversity in grammar and language style, ensuring that samples in the Biology-Instructions dataset have different question-answer styles. For examples of question-answer template pairs, please refer to Table~\ref{table:templates}.

\textbf{Reasoning} Although previous studies~\citep{richard2024chatnt, liu2024Genomic_Large_Language_Models, lv2024prollama} have demonstrated that large-scale primary instruction-following datasets can teach LLMs to answer biological sequence-related questions, they often fail to fully harness the powerful language abilities of LLMs, as they focus primarily on basic language patterns. In other words, they fail to leverage the powerful conversational abilities of these models to form natural and fluent dialogues or to utilize reasoning to improve the validity of the output results. To address this limitation, we design a prompt that requires powerful closed-source LLMs to reformulate answers for a subset of the validation set of Biology-Instructions and provide polished answers ready for end-users to read and understand, based on given questions and original answers. We encourage the model to first thoroughly analyze the sequence and the question, and then generate a well-reasoned, refined answer grounded in the preceding analysis and logical reasoning.

\subsection{Biology-Instructions Construction}

\begin{figure*}[ht]
\begin{center}
\includegraphics[width=\textwidth]{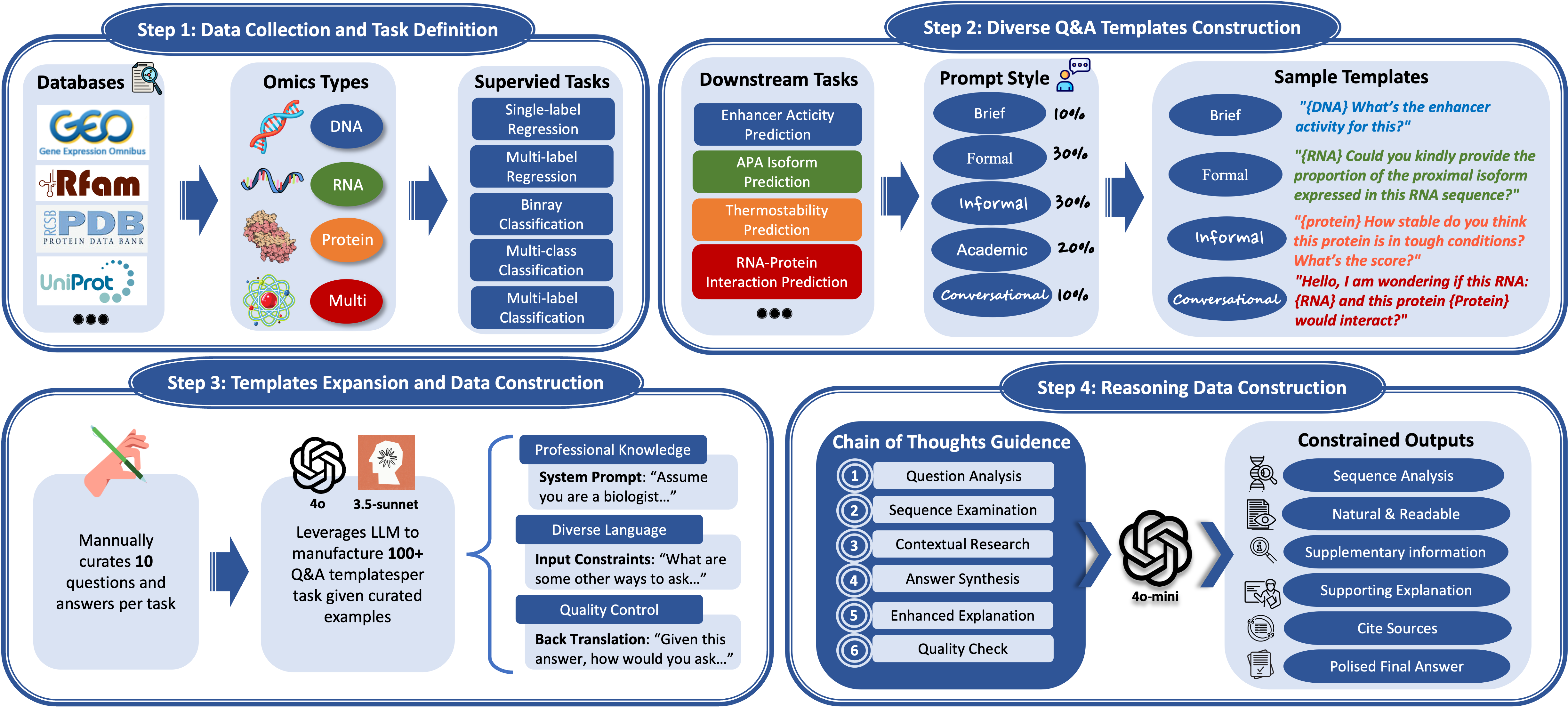}
\end{center}
\vspace{-3mm}
\caption{Overview of our data construction pipeline: Step1: Collect data from primary databases and categorize downstream tasks; Step2: Create diverse instruction prompts based on tasks; Step3: Use LLMs to enhance dataset quality; Step4: Follow key principles for reasoning data construction.}
\label{fig: Biology-Instructions Construction.}
\vspace{-3mm}
\end{figure*}

\subsubsection{Tasks}

As presented in Figure~\ref{fig: dataset distribution} (train-test split information is detailed in Table~\ref{Appendix:tasks}), the Biology-Instructions dataset comprises 21 tasks: 6 DNA tasks, 6 RNA tasks, 5 protein tasks, and 4 multi-molecule tasks. 
When considering the number of input sequences, there are 4 multi-molecule interaction tasks and 17 single-molecule tasks. 
Tasks were sourced from high-impact literature, journals, and competitions, ensuring coverage of biologically critical aspects in structure, function, and engineering across DNA, RNA, proteins, and their interactions.
We focus on predictive sequence-understanding tasks, leaving generative applications, such as sequence design, for future research. 
To the best of our knowledge, Biology-Instructions is the first instruction dataset to include multi-omics tasks and multi-molecule interaction tasks. 
\textbf{Task definitions are provided in Appendix~\ref{Appendix:section tasks}; the evaluation metric for each task is described in Appendix~\ref{Appendix:metrics}; and the input-output formats are detailed in Table~\ref{table: hint} and Table~\ref{table:templates}.}

\subsubsection{Templates}
To convert the original classification and regression task dataset into an instruction tuning dataset, we employ question-answer templates to integrate the data. The primary objective of creating these templates is to teach the model how to follow biological instructions and complete tasks without overfitting to specific language patterns. To achieve this, we prioritize diversity in language styles, expressive tones and lengths during the template construction process. We manually constructed 10 question templates and 10 answer templates for each task, covering various styles including, but not limited to, request, concise, informal, and academic styles. Then, we used GPT-4o and Claude-3.5-sonnet to expand the templates. Depending on the data volume for each task, we included 100 to 300 question templates and 100 to 300 answer templates. Ultimately, each task resulted in 10,000 to 100,000 question-answer template pairs. Since biological sequences are generally much longer than natural language prompts, we place the biological sequence at the beginning of question templates for each biological sequence task for non-interaction tasks. This approach helps prevent the prompts from being overwhelmed by the lengthy biological sequences, ensuring that the model can accurately understand the question and complete the task. 
Figure~\ref{fig: instruction prompts} provides examples of the instruction prompts constructed for each type of omics, illustrating the diversity and structure of the templates used in the dataset.

\subsubsection{Reasoning Data Construction}
Similar to the data construction method used by LlaVA~\citep{liu2024llava}. For a biological sequence \(X_s\) and its related question \(X_q\), simple answer \(Y_s\), we prompt GPT-4o-Mini to construct an optimized answer \(Y_o\) based on the given information. Generally, the instruction data were transformed to the format USER:\(X_s, X_q \) ASSISTANT:\(Y_o\). 

In the system prompt used for GPT-4o-Mini, as shown in Figure \ref{fig: stage3_prompt}, we emphasized the following key points to ensure the production of high-quality data: (1) first understand the provided biological sequence and the question; (2) analyze the biological sequence at the nucleotide or amino acid level, aiming to extract question-related information from the sequence; (3) refine the answer based on the previous analysis, including a rational explanation and a chain of thought approach, especially for complex questions; (4) list any relevant knowledge and information from reliable sources, and cite these sources appropriately; (5) return the polished answer in an end-to-end style, excluding any information from the standard answer and task hint. By following this approach, we gathered 8000 final AI-polished training data points without two multi-molecule tasks: antibody-antigen neutralization and RNA-protein interaction prediction, to study transfer learning for reasoning capability.
Figure \ref{fig: Biology-Instructions Construction.} provides an overview of the complete construction process for Biology-Instructions, including the data collection, template construction, and reasoning data construction stages.

\subsection{Evaluation Pipeline and Metrics}

Our evaluation framework is designed for robustly assessing model performance on the Biology-Instructions dataset. Tasks, irrespective of their associated omics domain, are categorized into five types: single-label regression, multi-label regression, binary classification, multi-class classification, and multi-label classification—each requiring tailored evaluation metrics to capture nuanced performance differences. The evaluation pipeline begins with preprocessing the models’ outputs, followed by grouping entries according to task type, and then computing task-specific metrics. For reporting, all metric values are scaled by 100 and rounded to two decimal places. Further details are provided in Appendix~\ref{Appendix:metrics}.

\begin{figure*}[t]
\begin{center}
\includegraphics[width=0.9\textwidth]{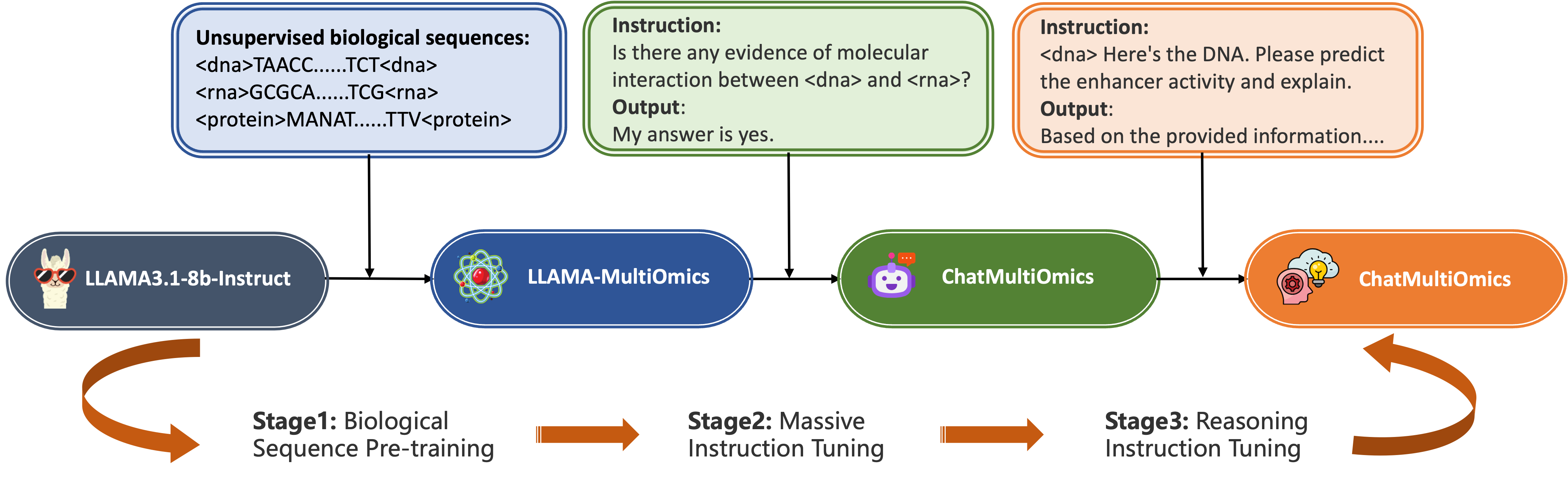}
\end{center}
\vspace{-3mm}
\caption{Overview of our three-stage training pipeline.}
\label{fig: training pipeline}
\vspace{-3mm}
\end{figure*}

\section{Model}

\label{headings}

As shown in Figure \ref{fig: training pipeline}, we train a model based on Llama3.1-8B-Instruct~\citep{dubey2024llama3} named ChatMultiOmics using multi-omics pre-training data and Biology-Instructions. In general, we perform a three-stage training paradigm to enhance the interactive, biology-related chat performance of the final biological assistant. \textbf{Please refer to Appendix~\ref{Appendix:C} for more training details.}

\subsection{Stage 1: Biological Sequences Continued Pre-training}

Although the memory savings facilitated by LoRA~\citep{devalal2018lora} are not that obvious when optimizer states are distributed across GPUs compared with training on a single GPU, LoRA can still significantly reduce training time by minimizing communication between data parallel ranks. However, directly applying LoRA to train a chat model on Biology-Instructions results in suboptimal performance on specific downstream tasks. Specifically, the model shows near-random performance in classification and regression tasks. As noted by \citet{limitation-of-it}, LoRA fine-tuning primarily leverages pre-trained knowledge to generate well-formed answers based on the output format learned from SFT data. We suspect that large-scale LoRA instruction tuning on biological sequence-related data suffers due to the lack of pre-training on biological sequence data, regardless of model architecture~\citep{ye2022b, ye2022efficient}, as evident from the baseline results. Therefore, continued pre-training of the model is essential for better performance. This involves teaching the model with biological sequences to enable it to understand the nature and functions of biological sequences. For this process, we utilized unlabeled human DNA data from the Genome Reference Consortium Human Genome (GRCh)~\citep{harrow2012gencode}, human non-coding RNA data from RNACentral~\citep{sweeney2019rnacentral}, and protein sequences from UniRef50~\citep{suzek2007uniref} during the first phase of pre-training. This initial pre-training served as a foundational warm-up to improve the model's comprehension across multi-omics biological sequences.

We employed \textbf{LoRA+}~\citep{hayou2024lora+} for all linear layers in the model, training on a continued pre-training dataset. LoRA+ demonstrates superior convergence compared to vanilla LoRA~\citep{hu2022lora} by increasing the learning rate of the weight B relative to the base learning rate for the weight A and other trainable parameters. \citet{hayou2024lora+} observed that setting the learning rate of weight B to 16 times that of weight A results in more effective convergence. However, our experiments revealed that while LoRA+ indeed improves convergence rates, applying a large learning rate multiplier can lead to instability during the continued pre-training process. Based on this observation, we opted for a more conservative learning rate multiplier of 4. We trained the normalization layers of the model alongside the LoRA parameters.

\begin{figure*}[ht]
\begin{center}
\includegraphics[width=\textwidth]{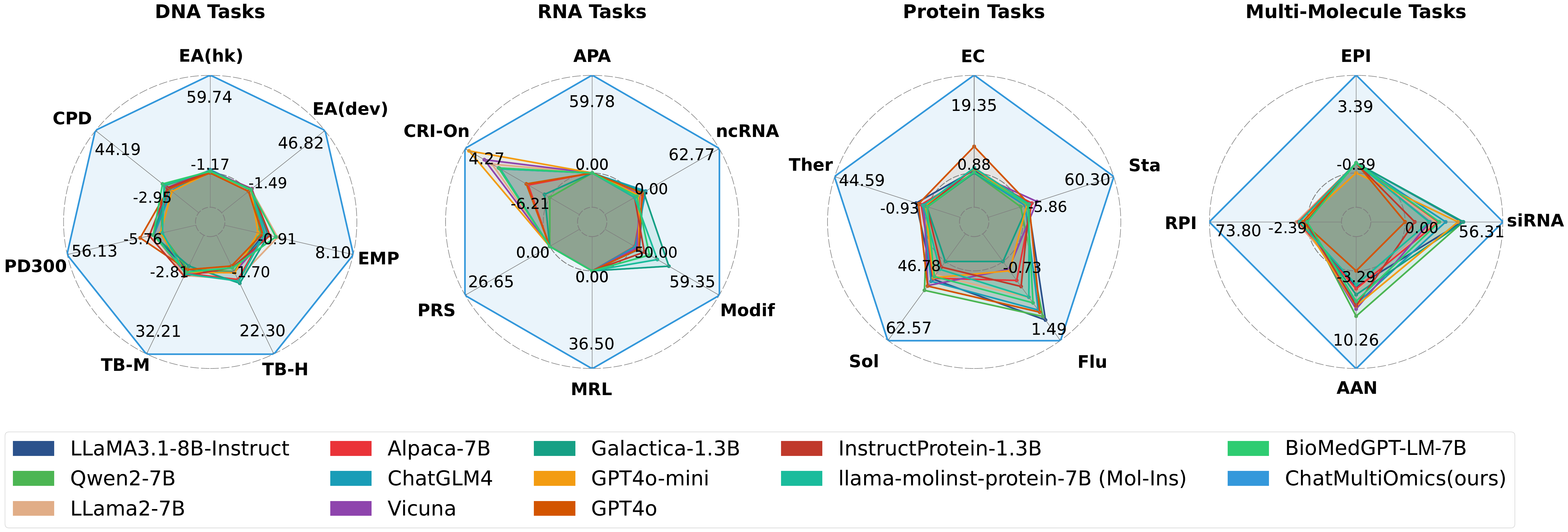}
\end{center}
\vspace{-5mm}
\caption{Radar plot comparing the performance of ChatMultiOmics with SOTA baselines on all 21 downstream tasks grouped by omics, including DNA, RNA, Protein, and Multi-molecule tasks.}
\label{fig: result-radar-plot}
\vspace{-3mm}
\end{figure*}

\subsection{Stage 2: Massive Instruction Tuning}
In Stage 2, we employ the Biology-Instructions dataset, excluding the reasoning sub-dataset. In the initial attempts of the training process, we find that the imbalance among tasks within the dataset can pose challenges for the model in distinguishing between different tasks. To mitigate this, we randomly select 30 percent of the training data and prepend a task label in the format "\texttt{[Classification/Regression:task\_name]}" at the beginning of each question. This method effectively aids the model in identifying different tasks and output objectives.

We use a system prompt \(P_{sc}\): "You are a knowledgeable and helpful biology assistant. Please answer my biology sequence-related questions clearly and concisely. For regression tasks, please return a number." This prompt helps the model to differentiate biology sequence-related tasks from other tasks. As illustrated in Figure \ref{fig: stage 2 traning sample}, we maintain the data format: SYSTEM:\(P_{sc}\) USER:\(X_s, X_q \) ASSISTANT:\(Y_o\) consistent with the chat template of Llama-3.1-8B-Instruct.

\subsection{Stage 3: Reasoning Instruction Tuning}
In Stage 3, we fine-tune the model using the reasoning sub-dataset from Biology-Instructions. To preserve the model's performance on classification and regression tasks, we additionally include 3,000 samples from the non-reasoning portion of the validation set, which are trained simultaneously.

To better control the behavior of the model, a more detailed system prompt \(P_{sd}\) was used for reasoning data: "You are a highly knowledgeable AI assistant specializing in biology, particularly in sequence-related topics. Your primary task is to provide clear, accurate, and comprehensive answers to biology questions. When analyzing and interpreting sequences, ensure to provide step-by-step explanations to make your responses natural and easy to understand. Engage with the user by asking clarifying questions if needed and offer detailed insights into the biological sequences." In this case, the format of training samples of reasoning data is transformed to SYSTEM:\(P_{sd}\) USER:\(X_s, X_q \) ASSISTANT:\(Y_o\).


\section{Results}
\label{results}

\begin{figure*}[ht!]
\begin{center}
\includegraphics[width=\textwidth]{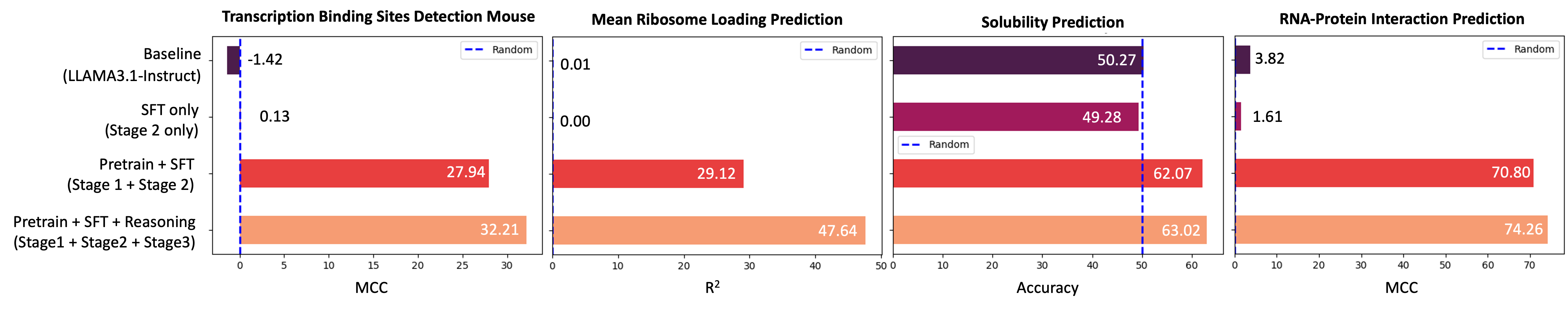}
\end{center}
\caption{Ablation studies showing the performance across different training stages. One downstream task from each omics type is selected for display.
Each bar color represents a specific training approach. The blue dashed line indicates the performance level expected by random chance for each task, based on the respective evaluation metric.
}
\label{fig7_ablation}
\end{figure*}

\subsection{Experimental Setups}
To evaluate the biological sequence understanding capabilities of current LLMs and determine if our method can enhance LLMs' performance, we compare ChatMultiOmics with various open-source general-purpose LLMs: Llama3.1-8B-Instruct~\citep{dubey2024llama3}, Llama2-7B-Chat~\citep{touvron2023llama2}, Alpaca-7B~\citep{taori2023alpaca}, Vicuna-v1.5-7B~\citep{chiang2023vicuna}, Qwen2-7B~\citep{bai2023qwen}, GLM4-9B-Chat~\citep{glm2024chatglm}, and Galactica-1.3b~\citep{taylor2022galactica}. Additionally, we include comparisons with SOTA closed-source LLMs: GPT-4o and GPT-4o-Mini. We also evaluate biology-specialized LLMs: InstructProtein-1.3B~\citep{wang2023instructprotein}, Llama-molinst-protein-7B~\citep{fang2023mol}, and BioMedGPT-LM-7B~\citep{zhang2023biomedgpt}. To ensure well-formed and quantifiable answers, we restrict the output format for all baselines and provide them with task information, enabling them to understand both what to output and how to format their output. The experimental results are visualized in Figure~\ref{fig: result-radar-plot}, showcasing the comparative performance of various LLMs across four types of datasets: DNA, RNA, protein, and multi-molecule interactions. \textbf{Due to space constraints, we present only the visual results here; the complete experimental results, including domain-specific baselines, are provided in tabular form in Appendix~\ref{Appendix.D} (Tables~\ref{table: dna result}--\ref{table: multi-molecule result}) for detailed numerical comparison.}

\subsection{Finding 1: General LLMs lack biological sequence understanding capability.}
To assess whether LLMs can effectively tackle tasks related to biological sequences, we conducted comprehensive experiments using both open-source and closed-source general-purpose LLMs. For open-source LLMs, we selected models of comparable size to our model, ChatMultiOmics. For closed-source LLMs, we evaluated SOTA models such as GPT-4o and its streamlined version, GPT-4o-mini. The results unequivocally demonstrate that all open-source LLMs of similar size to ChatMultiOmics fail to surpass average performance levels. Similarly, the closed-source LLMs, GPT-4o and GPT-4o-mini, exhibit performance on par with the open-source models. 

Notably, models within the same series but of different versions—such as Llama2-7B-Chat and Llama3.1-8B-Instruct—or of distinct sizes—such as GPT-4o and GPT-4o-mini—exhibit comparable performance on tasks involving biological sequences. These findings suggest that the language capabilities of these models do not directly correlate with their biological sequence understanding capability. This implies that natural language performance does not determine the effectiveness of these models in biological sequence understanding tasks, indicating a significant lack of pre-training on biological sequence knowledge. Despite their extensive text-based biological knowledge, LLMs struggle to connect this knowledge with biological sequences and are unable to analyze sequences at the molecular level effectively.

\subsection{Finding 2: Current Biology-specialized LLMs fail on multi-omics tasks.}
Biology-specified LLMs have demonstrated remarkable performance on a variety of tasks. For instance, the Llama-molinst-protein-7B model excels in five key areas of protein understanding, including the prediction of catalytic activity, protein design, protein function prediction, and others. Despite these impressive achievements, these methods exhibit limitations. Notably, they lack transfer learning capabilities across multi-omics tasks and fail to outperform general-purpose baselines even in single-omics tasks, and in some cases, struggle to follow instructions correctly. This indicates that while specialized LLMs are highly effective within their specific domains, their applicability and efficiency in broader, more integrative biological studies remain constrained.

\begin{table*}[!ht]
\caption{Performance comparison on protein-related tasks (ROUGE-L scores, higher is better).}
\label{table:mol_instructions}
\centering
\resizebox{\textwidth}{!}{%
\begin{tabular}{@{}lcccc@{}}
\toprule
\textbf{Model} & \textbf{Protein Function} & \textbf{Functional Description} & \textbf{Catalytic Activity} & \textbf{Domain/Motif} \\
\midrule
Training Set Size & 116,458 & 88,259 & 54,259 & 46,028 \\
\midrule
Alpaca & 0.20 & 0.10 & 0.23 & 0.12 \\
Baize & 0.20 & 0.15 & 0.22 & 0.13 \\
ChatGLM & 0.15 & 0.14 & 0.13 & 0.10 \\
LLaMa & 0.12 & 0.12 & 0.13 & 0.09 \\
Vicuna & 0.15 & 0.14 & 0.16 & 0.12 \\
Galactica & 0.07 & 0.08 & 0.08 & 0.06 \\
Llama-molinst-protein-7B & 0.43 & 0.44 & 0.52 & 0.46 \\
\textbf{ChatMultiOmics} & \textbf{0.36} & \textbf{0.38} & \textbf{0.60} & \textbf{0.44} \\
\bottomrule
\end{tabular}%
}
\end{table*}

\subsection{Finding 3: Continued pre-training boosts fine-tuning performance.}
Previous studies have utilized LoRA~\citep{fang2023mol, lv2024prollama} for model training. However, our experimental findings suggest that employing LoRA to fine-tune models on Biology-Instructions does not result in expected performance enhancements. For LoRA fine-tuning, the quality and quantity of the pre-training on related knowledge appear to be critical factors for achieving good results, as indirectly proved by the experimental setup in~\citep{fang2023mol}, where full fine-tuning was applied to protein-related tasks, and LoRA was used for the fine-tuning of other tasks, alongside the near-random performance of the baselines on biological-sequence understanding tasks. After continued pre-training on multi-omics sequences, LoRA fine-tuning on Biology-Instructions does help the model leverage the intrinsic relationships and dependencies from pre-trained knowledge. The results of the second stage significantly surpass those of instruction tuning without continued pre-training, as shown in Figure~\ref{fig7_ablation}.

\subsection{Finding 4: Fine-tuning on the reasoning dataset further enhances performance.}
We hypothesize that the model's performance can be enhanced by incorporating task information and reasoning steps, which can aid the model in better understanding the task and consequently lead to improved results. We tested the third-stage model using the system prompt \(P_{sc}\) to facilitate results computation. The results indicate that for most tasks, performance was enhanced. However, for some regression tasks, the performance was slightly adversely affected by the third-stage training. 

Furthermore, when the reasoning system prompt \(P_{sd}\) was used, the model demonstrated excellent reasoning capabilities and extended its performance to untrained tasks, such as antibody-antigen neutralization and RNA-protein interaction prediction, as illustrated in Figure~\ref{fig: fig1_ours_examples} (b).

\subsection{Finding 5: ChatMultiOmics generalizes to out-of-domain biological sequence tasks.}
Although ChatMultiOmics is primarily trained on a curated set of prediction and regression tasks, we further evaluate its ability to generalize to out-of-domain biological sequence tasks. To this end, we conduct a comprehensive benchmark across all protein-related tasks included in Mol-Instructions~\citep{fang2023mol}. For each task, the model is fine-tuned using only 1,000 samples for prompt-answer format alignment and then evaluated on the respective test sets. As shown in Table~\ref{table:mol_instructions}, ChatMultiOmics achieves highly competitive performance across all four tasks—Protein Function, Functional Description, Catalytic Activity, and Domain/Motif Prediction—despite the limited task-specific tuning. These results demonstrate the strong generalization capability of ChatMultiOmics to diverse protein-related tasks beyond its original training domain.

\section{Conclusion}

We introduce Biology-Instructions, the first large-scale, multi-omics instruction-tuning dataset focused on biological sequences. Biology-Instructions bridges the gap between large language models (LLMs) and complex biological tasks by encompassing 21 diverse tasks involving DNA, RNA, proteins, and multi-molecule interactions, spanning both single-sequence analysis and interaction-based reasoning. By incorporating reasoning-oriented instructions, Biology-Instructions enables LLMs to handle complex biological tasks with greater versatility while preserving conversational fluency.
Our evaluation reveals that SOTA LLMs such as GPT-4 struggle with biological sequence-related tasks without specialized training. Using Biology-Instructions for instruction tuning, we demonstrate significant improvements, proving its value in enhancing LLMs for multi-omics sequence analysis. We also develop ChatMultiOmics, along with a three-stage training pipeline: continued pre-training, massive instruction tuning, and reasoning instruction tuning. This pipeline yields significant performance improvements, offering an effective and scalable approach for training biological LLMs.

\section{Limitations} 
While Biology-Instructions is a significant advancement, it still has areas for improvement. The dataset covers primarily the predictive tasks. Future versions of this work should include generative tasks, such as designing novel protein sequences, which could greatly enhance its utility in protein engineering.

ChatMultiOmics demonstrates promising reasoning capabilities, yet further enhancements are needed to make its outputs more practical and reliable. To enhance model performance, we could use hybrid architectures that combine specialized biological tokenizers or encoders with LLMs and even refine the model architecture~\citep{10950128}. This could reduce information loss during the tokenization of biological sequences. 

Integrating structural data, such as 3D molecular coordinates, could improve the model's ability to capture functional implications of molecular structures.
Incorporating multi-hop data could be another potential enhancement for the model to reason over interconnected biological datasets and capture more intricate relationships across multiple omics layers.

Future efforts should also expand evaluation metrics beyond accuracy to include interpretability, robustness, and computational efficiency, offering a more holistic view of model performance.

Long-chain-of-thought reasoning models have demonstrated strong performance across a variety of complex tasks, such as advanced mathematical reasoning. These models also hold significant potential for excelling in biological sequence-related tasks as they can further bridge the gap between natural language and biological language. However, benchmarking long chain-of-thought reasoning models and developing more powerful models through such approaches is a comprehensive and labor-intensive endeavor, which we leave for future research. Future work could involve benchmarking various long-chain-of-thought reasoning models—such as DeepSeek-R1~\citep{guo2024deepseek} and QwQ-32B~\citep{qwq32b} on biology-sequence-related tasks using the Biology-Instructions dataset. 

Additionally, future works may explore constructing a long-chain-of-thought-enhanced version of Biology-Instructions through rejection sampling techniques. Furthermore, by leveraging this enhanced dataset alongside supervised fine-tuning or reinforcement learning methods, future studies can aim to improve the biological sequence understanding capabilities of LLMs.
\section*{Acknowledgments}
This work was supported by the Shanghai Artificial Intelligence Laboratory and a locally commissioned task from the Shanghai Municipal Government.

\bibliography{acl_latex}

\begin{thebibliography}{93}
\providecommand{\natexlab}[1]{#1}

\bibitem[{htm()}]{html1}

\newblock \url{http://www.jaist.ac.jp/~tran/nucleosome/members.html}.
\newblock Accessed: 2024-09-26.

\bibitem[{Achiam et~al.(2023)Achiam, Adler, Agarwal, Ahmad, Akkaya, Aleman, Almeida, Altenschmidt, Altman, Anadkat et~al.}]{achiam2023gpt}
Josh Achiam, Steven Adler, Sandhini Agarwal, Lama Ahmad, Ilge Akkaya, Florencia~Leoni Aleman, Diogo Almeida, Janko Altenschmidt, Sam Altman, Shyamal Anadkat, et~al. 2023.
\newblock Gpt-4 technical report.
\newblock \emph{arXiv preprint arXiv:2303.08774}.

\bibitem[{Angenent-Mari et~al.(2020{\natexlab{a}})Angenent-Mari, Garruss, Soenksen, Church, and Collins}]{angenent2020deep}
Nicolaas~M Angenent-Mari, Alexander~S Garruss, Luis~R Soenksen, George Church, and James~J Collins. 2020{\natexlab{a}}.
\newblock A deep learning approach to programmable rna switches.
\newblock \emph{Nature communications}, 11(1):5057.

\bibitem[{Angenent-Mari et~al.(2020{\natexlab{b}})Angenent-Mari, Garruss, Soenksen, Church, and Collins}]{Mari2020programmable}
Nicolaas~M. Angenent-Mari, Alexander~S. Garruss, Luis~R. Soenksen, George Church, and James~J. Collins. 2020{\natexlab{b}}.
\newblock \href {https://doi.org/10.1038/s41467-020-18677-1} {A deep learning approach to programmable rna switches}.
\newblock \emph{Nature Communications}, 11(1).

\bibitem[{Bai et~al.(2023)Bai, Bai, Chu, Cui, Dang, Deng, Fan, Ge, Han, Huang et~al.}]{bai2023qwen}
Jinze Bai, Shuai Bai, Yunfei Chu, Zeyu Cui, Kai Dang, Xiaodong Deng, Yang Fan, Wenbin Ge, Yu~Han, Fei Huang, et~al. 2023.
\newblock Qwen technical report.
\newblock \emph{arXiv preprint arXiv:2309.16609}.

\bibitem[{Bengio and LeCun(2007)}]{Bengio+chapter2007}
Yoshua Bengio and Yann LeCun. 2007.
\newblock Scaling learning algorithms towards {AI}.
\newblock In \emph{Large Scale Kernel Machines}. MIT Press.

\bibitem[{Bogard et~al.(2019{\natexlab{a}})Bogard, Linder, Rosenberg, and Seelig}]{bogard2019deep}
Nicholas Bogard, Johannes Linder, Alexander~B Rosenberg, and Georg Seelig. 2019{\natexlab{a}}.
\newblock A deep neural network for predicting and engineering alternative polyadenylation.
\newblock \emph{Cell}, 178(1):91--106.

\bibitem[{Bogard et~al.(2019{\natexlab{b}})Bogard, Linder, Rosenberg, and Seelig}]{bogard2019aparent}
Nicholas Bogard, Johannes Linder, Alexander~B Rosenberg, and Georg Seelig. 2019{\natexlab{b}}.
\newblock A deep neural network for predicting and engineering alternative polyadenylation.
\newblock \emph{Cell}, 178(1):91--106.

\bibitem[{Chen and Gong(2022)}]{chen2022hotprotein}
Tianlong Chen and Chengyue Gong. 2022.
\newblock Hotprotein: A novel framework for protein thermostability prediction and editing.
\newblock \emph{NeurIPS 2022}.

\bibitem[{Chiang et~al.(2023)Chiang, Li, Lin, Sheng, Wu, Zhang, Zheng, Zhuang, Zhuang, Gonzalez et~al.}]{chiang2023vicuna}
Wei-Lin Chiang, Zhuohan Li, Zi~Lin, Ying Sheng, Zhanghao Wu, Hao Zhang, Lianmin Zheng, Siyuan Zhuang, Yonghao Zhuang, Joseph~E Gonzalez, et~al. 2023.
\newblock Vicuna: An open-source chatbot impressing gpt-4 with 90\%* chatgpt quality.
\newblock \emph{See https://vicuna. lmsys. org (accessed 14 April 2023)}, 2(3):6.

\bibitem[{Chuai et~al.(2018)Chuai, Ma, Yan, Chen, Hong, Xue, Zhou, Zhu, Chen, Duan et~al.}]{chuai2018deepcrispr}
Guohui Chuai, Hanhui Ma, Jifang Yan, Ming Chen, Nanfang Hong, Dongyu Xue, Chi Zhou, Chenyu Zhu, Ke~Chen, Bin Duan, et~al. 2018.
\newblock Deepcrispr: optimized crispr guide rna design by deep learning.
\newblock \emph{Genome biology}, 19:1--18.

\bibitem[{Crick(1970)}]{crick1970central}
Francis Crick. 1970.
\newblock Central dogma of molecular biology.
\newblock \emph{Nature}, 227(5258):561--563.

\bibitem[{Dallago et~al.(2021)Dallago, Mou, Johnston, Wittmann, Bhattacharya, Goldman, Madani, and Yang}]{dallago2021flip}
Christian Dallago, Jody Mou, Kadina~E Johnston, Bruce~J Wittmann, Nicholas Bhattacharya, Samuel Goldman, Ali Madani, and Kevin~K Yang. 2021.
\newblock Flip: Benchmark tasks in fitness landscape inference for proteins.
\newblock \emph{bioRxiv}, pages 2021--11.

\bibitem[{Dao(2023)}]{dao2023flashattention2}
Tri Dao. 2023.
\newblock Flashattention-2: Faster attention with better parallelism and work partitioning.
\newblock \emph{arXiv preprint arXiv:2307.08691}.

\bibitem[{Dao et~al.(2022)Dao, Fu, Ermon, Rudra, and R{\'e}}]{dao2022flashattention}
Tri Dao, Dan Fu, Stefano Ermon, Atri Rudra, and Christopher R{\'e}. 2022.
\newblock Flashattention: Fast and memory-efficient exact attention with io-awareness.
\newblock \emph{Advances in Neural Information Processing Systems}, 35:16344--16359.

\bibitem[{de~Almeida et~al.(2022)de~Almeida, Reiter, Pagani, and Stark}]{de2022deepstarr}
Bernardo~P de~Almeida, Franziska Reiter, Michaela Pagani, and Alexander Stark. 2022.
\newblock Deepstarr predicts enhancer activity from dna sequence and enables the de novo design of synthetic enhancers.
\newblock \emph{Nature genetics}, 54(5):613--624.

\bibitem[{Devalal and Karthikeyan(2018)}]{devalal2018lora}
Shilpa Devalal and A~Karthikeyan. 2018.
\newblock Lora technology-an overview.
\newblock In \emph{2018 second international conference on electronics, communication and aerospace technology (ICECA)}, pages 284--290. IEEE.

\bibitem[{Devlin(2018)}]{devlin2018bert}
Jacob Devlin. 2018.
\newblock Bert: Pre-training of deep bidirectional transformers for language understanding.
\newblock \emph{arXiv preprint arXiv:1810.04805}.

\bibitem[{Dubey et~al.(2024)Dubey, Jauhri, Pandey, Kadian, Al-Dahle, Letman, Mathur, Schelten, Yang, Fan et~al.}]{dubey2024llama3}
Abhimanyu Dubey, Abhinav Jauhri, Abhinav Pandey, Abhishek Kadian, Ahmad Al-Dahle, Aiesha Letman, Akhil Mathur, Alan Schelten, Amy Yang, Angela Fan, et~al. 2024.
\newblock The llama 3 herd of models.
\newblock \emph{arXiv preprint arXiv:2407.21783}.

\bibitem[{Emilsson et~al.(2008)Emilsson, Thorleifsson, Zhang, Leonardson, Zink, Zhu, Carlson, Helgason, Walters, Gunnarsdottir et~al.}]{emilsson2008genetics}
Valur Emilsson, Gudmar Thorleifsson, Bin Zhang, Amy~S Leonardson, Florian Zink, Jun Zhu, Sonia Carlson, Agnar Helgason, G~Bragi Walters, Steinunn Gunnarsdottir, et~al. 2008.
\newblock Genetics of gene expression and its effect on disease.
\newblock \emph{Nature}, 452(7186):423--428.

\bibitem[{Fang et~al.(2023)Fang, Liang, Zhang, Liu, Huang, Chen, Fan, and Chen}]{fang2023mol}
Yin Fang, Xiaozhuan Liang, Ningyu Zhang, Kangwei Liu, Rui Huang, Zhuo Chen, Xiaohui Fan, and Huajun Chen. 2023.
\newblock Mol-instructions: A large-scale biomolecular instruction dataset for large language models.
\newblock \emph{arXiv preprint arXiv:2306.08018}.

\bibitem[{Fiannaca et~al.(2017)Fiannaca, La~Rosa, La~Paglia, Rizzo, and Urso}]{fiannaca2017nrc}
Antonino Fiannaca, Massimo La~Rosa, Laura La~Paglia, Riccardo Rizzo, and Alfonso Urso. 2017.
\newblock nrc: non-coding rna classifier based on structural features.
\newblock \emph{BioData mining}, 10:1--18.

\bibitem[{Ghosh et~al.(2024)Ghosh, Evuru, Kumar, Ramaneswaran, Aneja, Jin, Duraiswami, and Manocha}]{limitation-of-it}
Sreyan Ghosh, Chandra Kiran~Reddy Evuru, Sonal Kumar, S~Ramaneswaran, Deepali Aneja, Zeyu Jin, Ramani Duraiswami, and Dinesh Manocha. 2024.
\newblock A closer look at the limitations of instruction tuning.
\newblock In \emph{Forty-first International Conference on Machine Learning}.

\bibitem[{Gligorijevi{\'c} et~al.(2021)Gligorijevi{\'c}, Renfrew, Kosciolek, Leman, Berenberg, Vatanen, Chandler, Taylor, Fisk, Vlamakis et~al.}]{gligorijevic2021structure}
Vladimir Gligorijevi{\'c}, P~Douglas Renfrew, Tomasz Kosciolek, Julia~Koehler Leman, Daniel Berenberg, Tommi Vatanen, Chris Chandler, Bryn~C Taylor, Ian~M Fisk, Hera Vlamakis, et~al. 2021.
\newblock Structure-based protein function prediction using graph convolutional networks.
\newblock \emph{Nature communications}, 12(1):3168.

\bibitem[{GLM et~al.(2024)GLM, Zeng, Xu, Wang, Zhang, Yin, Rojas, Feng, Zhao, Lai et~al.}]{glm2024chatglm}
Team GLM, Aohan Zeng, Bin Xu, Bowen Wang, Chenhui Zhang, Da~Yin, Diego Rojas, Guanyu Feng, Hanlin Zhao, Hanyu Lai, et~al. 2024.
\newblock Chatglm: A family of large language models from glm-130b to glm-4 all tools.
\newblock \emph{arXiv preprint arXiv:2406.12793}.

\bibitem[{Goodfellow et~al.(2016)Goodfellow, Bengio, Courville, and Bengio}]{goodfellow2016deep}
Ian Goodfellow, Yoshua Bengio, Aaron Courville, and Yoshua Bengio. 2016.
\newblock \emph{Deep learning}, volume~1.
\newblock MIT Press.

\bibitem[{Guo et~al.(2025)Guo, Yang, Zhang, Song, Zhang, Xu, Zhu, Ma, Wang, Bi et~al.}]{guo2025deepseek}
Daya Guo, Dejian Yang, Haowei Zhang, Junxiao Song, Ruoyu Zhang, Runxin Xu, Qihao Zhu, Shirong Ma, Peiyi Wang, Xiao Bi, et~al. 2025.
\newblock Deepseek-r1: Incentivizing reasoning capability in llms via reinforcement learning.
\newblock \emph{arXiv preprint arXiv:2501.12948}.

\bibitem[{Guo et~al.(2024)Guo, Zhu, Yang, Xie, Dong, Zhang, Chen, Bi, Wu, Li et~al.}]{guo2024deepseek}
Daya Guo, Qihao Zhu, Dejian Yang, Zhenda Xie, Kai Dong, Wentao Zhang, Guanting Chen, Xiao Bi, Yu~Wu, YK~Li, et~al. 2024.
\newblock Deepseek-coder: When the large language model meets programming--the rise of code intelligence.
\newblock \emph{arXiv preprint arXiv:2401.14196}.

\bibitem[{Han and Zhang(2023)}]{han2023ncrpi}
Yong Han and Shao-Wu Zhang. 2023.
\newblock ncrpi-lgat: Prediction of ncrna-protein interactions with line graph attention network framework.
\newblock \emph{Computational and Structural Biotechnology Journal}, 21:2286--2295.

\bibitem[{Harrow et~al.(2012)Harrow, Frankish, Gonzalez, Tapanari, Diekhans, Kokocinski, Aken, Barrell, Zadissa, Searle et~al.}]{harrow2012gencode}
Jennifer Harrow, Adam Frankish, Jose~M Gonzalez, Electra Tapanari, Mark Diekhans, Felix Kokocinski, Bronwen~L Aken, Daniel Barrell, Amonida Zadissa, Stephen Searle, et~al. 2012.
\newblock Gencode: the reference human genome annotation for the encode project.
\newblock \emph{Genome research}, 22(9):1760--1774.

\bibitem[{Hayou et~al.(2024)Hayou, Ghosh, and Yu}]{hayou2024lora+}
Soufiane Hayou, Nikhil Ghosh, and Bin Yu. 2024.
\newblock Lora+: Efficient low rank adaptation of large models.
\newblock \emph{arXiv preprint arXiv:2402.12354}.

\bibitem[{Hinton et~al.(2006)Hinton, Osindero, and Teh}]{Hinton06}
Geoffrey~E. Hinton, Simon Osindero, and Yee~Whye Teh. 2006.
\newblock A fast learning algorithm for deep belief nets.
\newblock \emph{Neural Computation}, 18:1527--1554.

\bibitem[{Hu et~al.(2022{\natexlab{a}})Hu, Shen, Wallis, Allen-Zhu, Li, Wang, Wang, Chen et~al.}]{hu2022lora}
Edward~J Hu, Yelong Shen, Phillip Wallis, Zeyuan Allen-Zhu, Yuanzhi Li, Shean Wang, Lu~Wang, Weizhu Chen, et~al. 2022{\natexlab{a}}.
\newblock Lora: Low-rank adaptation of large language models.
\newblock \emph{ICLR}, 1(2):3.

\bibitem[{Hu et~al.(2022{\natexlab{b}})Hu, Yuan, Yang, Ju, Su, Wang, Yang, and Ding}]{hu2022exploring}
Mingyang Hu, Fajie Yuan, Kevin Yang, Fusong Ju, Jin Su, Hui Wang, Fei Yang, and Qiuyang Ding. 2022{\natexlab{b}}.
\newblock Exploring evolution-aware \&-free protein language models as protein function predictors.
\newblock \emph{Advances in Neural Information Processing Systems}, 35:38873--38884.

\bibitem[{Jacobs et~al.(2023)Jacobs, Tanaka, Zhang, Zhang, Song, Rajbhandari, and He}]{jacobs2023deepspeed}
Sam~Ade Jacobs, Masahiro Tanaka, Chengming Zhang, Minjia Zhang, Leon Song, Samyam Rajbhandari, and Yuxiong He. 2023.
\newblock Deepspeed ulysses: System optimizations for enabling training of extreme long sequence transformer models.
\newblock \emph{arXiv preprint arXiv:2309.14509}.

\bibitem[{Kalajdzievski(2023)}]{kalajdzievski2023rslora}
Damjan Kalajdzievski. 2023.
\newblock A rank stabilization scaling factor for fine-tuning with lora.
\newblock \emph{arXiv preprint arXiv:2312.03732}.

\bibitem[{Khurana et~al.(2018)Khurana, Rawi, Kunji, Chuang, Bensmail, and Mall}]{khurana2018deepsol}
Sameer Khurana, Reda Rawi, Khalid Kunji, Gwo-Yu Chuang, Halima Bensmail, and Raghvendra Mall. 2018.
\newblock Deepsol: a deep learning framework for sequence-based protein solubility prediction.
\newblock \emph{Bioinformatics}, 34(15):2605--2613.

\bibitem[{Korthikanti et~al.(2023)Korthikanti, Casper, Lym, McAfee, Andersch, Shoeybi, and Catanzaro}]{korthikanti2023reducing}
Vijay~Anand Korthikanti, Jared Casper, Sangkug Lym, Lawrence McAfee, Michael Andersch, Mohammad Shoeybi, and Bryan Catanzaro. 2023.
\newblock Reducing activation recomputation in large transformer models.
\newblock \emph{Proceedings of Machine Learning and Systems}, 5:341--353.

\bibitem[{Li et~al.(2024)Li, Wong, Zhang, Usuyama, Liu, Yang, Naumann, Poon, and Gao}]{li2024llavamed}
Chunyuan Li, Cliff Wong, Sheng Zhang, Naoto Usuyama, Haotian Liu, Jianwei Yang, Tristan Naumann, Hoifung Poon, and Jianfeng Gao. 2024.
\newblock Llava-med: Training a large language-and-vision assistant for biomedicine in one day.
\newblock \emph{Advances in Neural Information Processing Systems}, 36.

\bibitem[{Li et~al.(2016)Li, You, Chen, Gui, and Nie}]{li2016highly}
Zheng-Wei Li, Zhu-Hong You, Xing Chen, Jie Gui, and Ru~Nie. 2016.
\newblock Highly accurate prediction of protein-protein interactions via incorporating evolutionary information and physicochemical characteristics.
\newblock \emph{International journal of molecular sciences}, 17(9):1396.

\bibitem[{Liu et~al.(2023)Liu, Zaharia, and Abbeel}]{liu2023ring}
Hao Liu, Matei Zaharia, and Pieter Abbeel. 2023.
\newblock Ring attention with blockwise transformers for near-infinite context.
\newblock \emph{arXiv preprint arXiv:2310.01889}.

\bibitem[{Liu et~al.(2024{\natexlab{a}})Liu, Li, Wu, and Lee}]{liu2024llava}
Haotian Liu, Chunyuan Li, Qingyang Wu, and Yong~Jae Lee. 2024{\natexlab{a}}.
\newblock Visual instruction tuning.
\newblock \emph{Advances in neural information processing systems}, 36.

\bibitem[{Liu et~al.(2024{\natexlab{b}})Liu, Zhou, Chen, Liu, Huo, and Han}]{liu2024Genomic_Large_Language_Models}
Huaqing Liu, Shuxian Zhou, Peiyi Chen, Jiahui Liu, Ku-Geng Huo, and Lanqing Han. 2024{\natexlab{b}}.
\newblock Exploring genomic large language models: Bridging the gap between natural language and gene sequences.
\newblock \emph{bioRxiv}, pages 2024--02.

\bibitem[{Liu et~al.(2021)Liu, Ji, Fu, Tam, Du, Yang, and Tang}]{liu2021ptuning}
Xiao Liu, Kaixuan Ji, Yicheng Fu, Weng~Lam Tam, Zhengxiao Du, Zhilin Yang, and Jie Tang. 2021.
\newblock P-tuning v2: Prompt tuning can be comparable to fine-tuning universally across scales and tasks.
\newblock \emph{arXiv preprint arXiv:2110.07602}.

\bibitem[{Lv et~al.(2024)Lv, Lin, Li, Liu, Cui, Chen, Yuan, and Tian}]{lv2024prollama}
Liuzhenghao Lv, Zongying Lin, Hao Li, Yuyang Liu, Jiaxi Cui, Calvin Yu-Chian Chen, Li~Yuan, and Yonghong Tian. 2024.
\newblock Prollama: A protein large language model for multi-task protein language processing.
\newblock \emph{arXiv preprint arXiv:2402.16445}.

\bibitem[{Mangrulkar et~al.(2022)Mangrulkar, Gugger, Debut, Belkada, Paul, and Bossan}]{peft}
Sourab Mangrulkar, Sylvain Gugger, Lysandre Debut, Younes Belkada, Sayak Paul, and Benjamin Bossan. 2022.
\newblock Peft: State-of-the-art parameter-efficient fine-tuning methods.
\newblock \url{https://github.com/huggingface/peft}.

\bibitem[{Marcotte et~al.(1999)Marcotte, Pellegrini, Ng, Rice, Yeates, and Eisenberg}]{marcotte1999detecting}
Edward~M Marcotte, Matteo Pellegrini, Ho-Leung Ng, Danny~W Rice, Todd~O Yeates, and David Eisenberg. 1999.
\newblock Detecting protein function and protein-protein interactions from genome sequences.
\newblock \emph{Science}, 285(5428):751--753.

\bibitem[{Mattick(2004)}]{mattick2004rna}
John~S Mattick. 2004.
\newblock Rna regulation: a new genetics?
\newblock \emph{Nature Reviews Genetics}, 5(4):316--323.

\bibitem[{Meier et~al.(2021)Meier, Rao, Verkuil, Liu, Sercu, and Rives}]{meier2021language}
Joshua Meier, Roshan Rao, Robert Verkuil, Jason Liu, Tom Sercu, and Alex Rives. 2021.
\newblock Language models enable zero-shot prediction of the effects of mutations on protein function.
\newblock \emph{Advances in neural information processing systems}, 34:29287--29303.

\bibitem[{Men et~al.(2024)Men, Xu, Wang, Zhang, Lin, Han, and Chen}]{men2024base}
Xin Men, Mingyu Xu, Bingning Wang, Qingyu Zhang, Hongyu Lin, Xianpei Han, and Weipeng Chen. 2024.
\newblock Base of rope bounds context length.
\newblock \emph{arXiv preprint arXiv:2405.14591}.

\bibitem[{Min et~al.(2021)Min, Ye, Liu, and Zeng}]{min2021predicting}
Xiaoping Min, Congmin Ye, Xiangrong Liu, and Xiangxiang Zeng. 2021.
\newblock Predicting enhancer-promoter interactions by deep learning and matching heuristic.
\newblock \emph{Briefings in Bioinformatics}, 22(4):bbaa254.

\bibitem[{Rajbhandari et~al.(2020)Rajbhandari, Rasley, Ruwase, and He}]{rajbhandari2020zero}
Samyam Rajbhandari, Jeff Rasley, Olatunji Ruwase, and Yuxiong He. 2020.
\newblock Zero: Memory optimizations toward training trillion parameter models.
\newblock In \emph{SC20: International Conference for High Performance Computing, Networking, Storage and Analysis}, pages 1--16. IEEE.

\bibitem[{Rao et~al.(2019)Rao, Bhattacharya, Thomas, Duan, Chen, Canny, Abbeel, and Song}]{rao2019evaluating}
Roshan Rao, Nicholas Bhattacharya, Neil Thomas, Yan Duan, Peter Chen, John Canny, Pieter Abbeel, and Yun Song. 2019.
\newblock Evaluating protein transfer learning with tape.
\newblock \emph{Advances in neural information processing systems}, 32.

\bibitem[{Ren et~al.(2024{\natexlab{a}})Ren, Chen, Qiao, Jing, Cai, Xu, Ye, Ma, Sun, Yan et~al.}]{ren2024beacon}
Yuchen Ren, Zhiyuan Chen, Lifeng Qiao, Hongtai Jing, Yuchen Cai, Sheng Xu, Peng Ye, Xinzhu Ma, Siqi Sun, Hongliang Yan, et~al. 2024{\natexlab{a}}.
\newblock Beacon: Benchmark for comprehensive rna tasks and language models.
\newblock \emph{arXiv preprint arXiv:2406.10391}.

\bibitem[{Ren et~al.(2024{\natexlab{b}})Ren, Han, Zhang, Tang, Bai, Cai, Qiao, Jiang, Yuan, Chen et~al.}]{ren2024comet}
Yuchen Ren, Wenwei Han, Qianyuan Zhang, Yining Tang, Weiqiang Bai, Yuchen Cai, Lifeng Qiao, Hao Jiang, Dong Yuan, Tao Chen, et~al. 2024{\natexlab{b}}.
\newblock Comet: Benchmark for comprehensive biological multi-omics evaluation tasks and language models.
\newblock \emph{arXiv preprint arXiv:2412.10347}.

\bibitem[{Richard et~al.(2024)Richard, de~Almeida, Dalla-Torre, Blum, Hexemer, Pandey, Laurent, Lopez, Laterre, Lang et~al.}]{richard2024chatnt}
Guillaume Richard, Bernardo~P de~Almeida, Hugo Dalla-Torre, Christopher Blum, Lorenz Hexemer, Priyanka Pandey, Stefan Laurent, Marie~P Lopez, Alexander Laterre, Maren Lang, et~al. 2024.
\newblock Chatnt: A multimodal conversational agent for dna, rna and protein tasks.
\newblock \emph{bioRxiv}, pages 2024--04.

\bibitem[{Rives et~al.(2021)Rives, Meier, Sercu, Goyal, Lin, Liu, Guo, Ott, Zitnick, Ma et~al.}]{rives2021esm}
Alexander Rives, Joshua Meier, Tom Sercu, Siddharth Goyal, Zeming Lin, Jason Liu, Demi Guo, Myle Ott, C~Lawrence Zitnick, Jerry Ma, et~al. 2021.
\newblock Biological structure and function emerge from scaling unsupervised learning to 250 million protein sequences.
\newblock \emph{Proceedings of the National Academy of Sciences}, 118(15):e2016239118.

\bibitem[{Rocklin et~al.(2017)Rocklin, Chidyausiku, Goreshnik, Ford, Houliston, Lemak, Carter, Ravichandran, Mulligan, Chevalier, Arrowsmith, and Baker}]{Rocklin2017Global}
Gabriel~J. Rocklin, Tamuka~M. Chidyausiku, Inna Goreshnik, Alex Ford, Scott Houliston, Alexander Lemak, Lauren Carter, Rashmi Ravichandran, Vikram~K. Mulligan, Aaron Chevalier, Cheryl~H. Arrowsmith, and David Baker. 2017.
\newblock \href {https://doi.org/10.1126/science.aan0693} {Global analysis of protein folding using massively parallel design, synthesis, and testing}.
\newblock \emph{Science}, 357(6347):168--175.

\bibitem[{Rossi et~al.(2019)Rossi, Monti, Bronstein, and Li{\`o}}]{rossi2019rnagcn}
Emanuele Rossi, Federico Monti, Michael Bronstein, and Pietro Li{\`o}. 2019.
\newblock ncrna classification with graph convolutional networks.
\newblock \emph{arXiv preprint arXiv:1905.06515}.

\bibitem[{SAIS(2020)}]{siRNAdata}
SAIS. 2020.
\newblock sirna data.
\newblock \url{http://competition.sais.com.cn/competitionDetail/532230/format}.
\newblock Accessed: 2024-05-26.

\bibitem[{Sample et~al.(2019{\natexlab{a}})Sample, Wang, Reid, Presnyak, McFadyen, Morris, and Seelig}]{sample2019human}
Paul~J Sample, Ban Wang, David~W Reid, Vlad Presnyak, Iain~J McFadyen, David~R Morris, and Georg Seelig. 2019{\natexlab{a}}.
\newblock Human 5' utr design and variant effect prediction from a massively parallel translation assay.
\newblock \emph{Nature biotechnology}, 37(7):803--809.

\bibitem[{Sample et~al.(2019{\natexlab{b}})Sample, Wang, Reid, Presnyak, McFadyen, Morris, and Seelig}]{sample2019MRL}
Paul~J Sample, Ban Wang, David~W Reid, Vlad Presnyak, Iain~J McFadyen, David~R Morris, and Georg Seelig. 2019{\natexlab{b}}.
\newblock Human 5' utr design and variant effect prediction from a massively parallel translation assay.
\newblock \emph{Nature biotechnology}, 37(7):803--809.

\bibitem[{Sarkisyan et~al.(2016)Sarkisyan, Bolotin, Meer, Usmanova, Mishin, Sharonov, Ivankov, Bozhanova, Baranov, Soylemez, Bogatyreva, Vlasov, Egorov, Logacheva, Kondrashov, Chudakov, Putintseva, Mamedov, Tawfik, Lukyanov, and Kondrashov}]{Sarkisyan2016Local}
Karen~S. Sarkisyan, Dmitry~A. Bolotin, Margarita~V. Meer, Dinara~R. Usmanova, Alexander~S. Mishin, George~V. Sharonov, Dmitry~N. Ivankov, Nina~G. Bozhanova, Mikhail~S. Baranov, Onuralp Soylemez, Natalya~S. Bogatyreva, Peter~K. Vlasov, Evgeny~S. Egorov, Maria~D. Logacheva, Alexey~S. Kondrashov, Dmitry~M. Chudakov, Ekaterina~V. Putintseva, Ilgar~Z. Mamedov, Dan~S. Tawfik, Konstantin~A. Lukyanov, and Fyodor~A. Kondrashov. 2016.
\newblock \href {https://doi.org/10.1038/nature17995} {Local fitness landscape of the green fluorescent protein}.
\newblock \emph{Nature}, 533(7603):397--401.

\bibitem[{Shanehsazzadeh et~al.(2020)Shanehsazzadeh, Belanger, and Dohan}]{shanehsazzadeh2020transfer}
Amir Shanehsazzadeh, David Belanger, and David Dohan. 2020.
\newblock Is transfer learning necessary for protein landscape prediction?
\newblock \emph{arXiv preprint arXiv:2011.03443}.

\bibitem[{Song et~al.(2021{\natexlab{a}})Song, Huang, Song, Chen, Song, Liu, Su, Magalh{\~a}es, Rigden, and Meng}]{song2021attention}
Zitao Song, Daiyun Huang, Bowen Song, Kunqi Chen, Yiyou Song, Gang Liu, Jionglong Su, Jo{\~a}o Pedro~de Magalh{\~a}es, Daniel~J Rigden, and Jia Meng. 2021{\natexlab{a}}.
\newblock Attention-based multi-label neural networks for integrated prediction and interpretation of twelve widely occurring rna modifications.
\newblock \emph{Nature communications}, 12(1):4011.

\bibitem[{Song et~al.(2021{\natexlab{b}})Song, Huang, Song, Chen, Song, Liu, Su, Magalh{\~a}es, Rigden, and Meng}]{song2021MultiRM}
Zitao Song, Daiyun Huang, Bowen Song, Kunqi Chen, Yiyou Song, Gang Liu, Jionglong Su, Jo{\~a}o Pedro~de Magalh{\~a}es, Daniel~J Rigden, and Jia Meng. 2021{\natexlab{b}}.
\newblock Attention-based multi-label neural networks for integrated prediction and interpretation of twelve widely occurring rna modifications.
\newblock \emph{Nature communications}, 12(1):4011.

\bibitem[{Su et~al.(2024{\natexlab{a}})Su, Ahmed, Lu, Pan, Bo, and Liu}]{su2024roformer}
Jianlin Su, Murtadha Ahmed, Yu~Lu, Shengfeng Pan, Wen Bo, and Yunfeng Liu. 2024{\natexlab{a}}.
\newblock Roformer: Enhanced transformer with rotary position embedding.
\newblock \emph{Neurocomputing}, 568:127063.

\bibitem[{Su et~al.(2024{\natexlab{b}})Su, Han, Zhou, Shan, Zhou, and Yuan}]{su2024saprot}
Jin Su, Chenchen Han, Yuyang Zhou, Junjie Shan, Xibin Zhou, and Fajie Yuan. 2024{\natexlab{b}}.
\newblock Saprot: Protein language modeling with structure-aware vocabulary.
\newblock In \emph{ICLR}.

\bibitem[{Suzek et~al.(2007)Suzek, Huang, McGarvey, Mazumder, and Wu}]{suzek2007uniref}
Baris~E Suzek, Hongzhan Huang, Peter McGarvey, Raja Mazumder, and Cathy~H Wu. 2007.
\newblock Uniref: comprehensive and non-redundant uniprot reference clusters.
\newblock \emph{Bioinformatics}, 23(10):1282--1288.

\bibitem[{Sweeney et~al.(2019)Sweeney, Petrov, Burkov, Finn, Bateman, Szymanski, Karlowski, Gorodkin, Seemann, Cannone et~al.}]{sweeney2019rnacentral}
Blake~A Sweeney, Anton~I Petrov, Boris Burkov, Robert~D Finn, Alex Bateman, Maciej Szymanski, Wojciech~M Karlowski, Jan Gorodkin, Stefan~E Seemann, Jamie~J Cannone, et~al. 2019.
\newblock Rnacentral: a hub of information for non-coding rna sequences.
\newblock \emph{NUCLEIC ACIDS RESEARCH}, 47(D1):D221--D229.

\bibitem[{Taori et~al.(2023)Taori, Gulrajani, Zhang, Dubois, Li, Guestrin, Liang, and Hashimoto}]{taori2023alpaca}
Rohan Taori, Ishaan Gulrajani, Tianyi Zhang, Yann Dubois, Xuechen Li, Carlos Guestrin, Percy Liang, and Tatsunori~B Hashimoto. 2023.
\newblock Alpaca: A strong, replicable instruction-following model.
\newblock \emph{Stanford Center for Research on Foundation Models. https://crfm. stanford. edu/2023/03/13/alpaca. html}, 3(6):7.

\bibitem[{Taylor et~al.(2022)Taylor, Kardas, Cucurull, Scialom, Hartshorn, Saravia, Poulton, Kerkez, and Stojnic}]{taylor2022galactica}
Ross Taylor, Marcin Kardas, Guillem Cucurull, Thomas Scialom, Anthony Hartshorn, Elvis Saravia, Andrew Poulton, Viktor Kerkez, and Robert Stojnic. 2022.
\newblock Galactica: A large language model for science.
\newblock \emph{arXiv preprint arXiv:2211.09085}.

\bibitem[{Team et~al.(2023)Team, Anil, Borgeaud, Wu, Alayrac, Yu, Soricut, Schalkwyk, Dai, Hauth et~al.}]{team2023gemini}
Gemini Team, Rohan Anil, Sebastian Borgeaud, Yonghui Wu, Jean-Baptiste Alayrac, Jiahui Yu, Radu Soricut, Johan Schalkwyk, Andrew~M Dai, Anja Hauth, et~al. 2023.
\newblock Gemini: a family of highly capable multimodal models.
\newblock \emph{arXiv preprint arXiv:2312.11805}.

\bibitem[{Team(2025)}]{qwq32b}
Qwen Team. 2025.
\newblock \href {https://qwenlm.github.io/blog/qwq-32b/} {Qwq-32b: Embracing the power of reinforcement learning}.

\bibitem[{Touvron et~al.(2023{\natexlab{a}})Touvron, Lavril, Izacard, Martinet, Lachaux, Lacroix, Rozi{\`e}re, Goyal, Hambro, Azhar et~al.}]{touvron2023llama1}
Hugo Touvron, Thibaut Lavril, Gautier Izacard, Xavier Martinet, Marie-Anne Lachaux, Timoth{\'e}e Lacroix, Baptiste Rozi{\`e}re, Naman Goyal, Eric Hambro, Faisal Azhar, et~al. 2023{\natexlab{a}}.
\newblock Llama: Open and efficient foundation language models.
\newblock \emph{arXiv preprint arXiv:2302.13971}.

\bibitem[{Touvron et~al.(2023{\natexlab{b}})Touvron, Martin, Stone, Albert, Almahairi, Babaei, Bashlykov, Batra, Bhargava, Bhosale et~al.}]{touvron2023llama2}
Hugo Touvron, Louis Martin, Kevin Stone, Peter Albert, Amjad Almahairi, Yasmine Babaei, Nikolay Bashlykov, Soumya Batra, Prajjwal Bhargava, Shruti Bhosale, et~al. 2023{\natexlab{b}}.
\newblock Llama 2: Open foundation and fine-tuned chat models.
\newblock \emph{arXiv preprint arXiv:2307.09288}.

\bibitem[{Vaswani(2017)}]{vaswani2017transformer}
A~Vaswani. 2017.
\newblock Attention is all you need.
\newblock \emph{Advances in Neural Information Processing Systems}.

\bibitem[{Wang et~al.(2023)Wang, Zhang, Ding, Qin, Zhuang, Li, and Chen}]{wang2023instructprotein}
Zeyuan Wang, Qiang Zhang, Keyan Ding, Ming Qin, Xiang Zhuang, Xiaotong Li, and Huajun Chen. 2023.
\newblock Instructprotein: Aligning human and protein language via knowledge instruction.
\newblock \emph{arXiv preprint arXiv:2310.03269}.

\bibitem[{Xu et~al.(2015)Xu, Xiao, Chen, Li, Meyer, Wu, Wu, Cong, Zhang, Liu et~al.}]{xu2015sequence}
Han Xu, Tengfei Xiao, Chen-Hao Chen, Wei Li, Clifford~A Meyer, Qiu Wu, Di~Wu, Le~Cong, Feng Zhang, Jun~S Liu, et~al. 2015.
\newblock Sequence determinants of improved crispr sgrna design.
\newblock \emph{Genome research}, 25(8):1147--1157.

\bibitem[{Xu et~al.(2024)Xu, Jiang, Niu, Deng, Poovendran, Choi, and Lin}]{xu2024magpie}
Zhangchen Xu, Fengqing Jiang, Luyao Niu, Yuntian Deng, Radha Poovendran, Yejin Choi, and Bill~Yuchen Lin. 2024.
\newblock Magpie: Alignment data synthesis from scratch by prompting aligned llms with nothing.
\newblock \emph{arXiv preprint arXiv:2406.08464}.

\bibitem[{yang Liu et~al.(2024)yang Liu, Wang, Yin, Molchanov, Wang, Cheng, and Chen}]{yang2024dora}
Shih yang Liu, Chien-Yi Wang, Hongxu Yin, Pavlo Molchanov, Yu-Chiang~Frank Wang, Kwang-Ting Cheng, and Min-Hung Chen. 2024.
\newblock Dora: Weight-decomposed low-rank adaptation.
\newblock \emph{ArXiv, abs/2402.09353}, 5.

\bibitem[{Ye et~al.(2025{\natexlab{a}})Ye, He, Tang, Li, Chen, Bai, and Ouyang}]{10950128}
Peng Ye, Tong He, Shengji Tang, Baopu Li, Tao Chen, Lei Bai, and Wanli Ouyang. 2025{\natexlab{a}}.
\newblock \href {https://doi.org/10.1109/TPAMI.2025.3558421} {Stimulative training++: Go beyond the performance limits of residual networks}.
\newblock \emph{IEEE Transactions on Pattern Analysis and Machine Intelligence}, 47(8):6463--6479.

\bibitem[{Ye et~al.(2025{\natexlab{b}})Ye, Huang, Shen, Chen, Huang, and Ouyang}]{ye2025dynamic}
Peng Ye, Chenyu Huang, Mingzhu Shen, Tao Chen, Yongqi Huang, and Wanli Ouyang. 2025{\natexlab{b}}.
\newblock Dynamic model merging with mixture of weights.
\newblock \emph{IEEE Transactions on Circuits and Systems for Video Technology}.

\bibitem[{Ye et~al.(2022{\natexlab{a}})Ye, Li, Chen, Fan, Mei, Lin, Zuo, Chi, and Ouyang}]{ye2022efficient}
Peng Ye, Baopu Li, Tao Chen, Jiayuan Fan, Zhen Mei, Chen Lin, Chongyan Zuo, Qinghua Chi, and Wanli Ouyang. 2022{\natexlab{a}}.
\newblock Efficient joint-dimensional search with solution space regularization for real-time semantic segmentation.
\newblock \emph{International Journal of Computer Vision}, 130(11):2674--2694.

\bibitem[{Ye et~al.(2022{\natexlab{b}})Ye, Li, Li, Chen, Fan, and Ouyang}]{ye2022b}
Peng Ye, Baopu Li, Yikang Li, Tao Chen, Jiayuan Fan, and Wanli Ouyang. 2022{\natexlab{b}}.
\newblock b-darts: Beta-decay regularization for differentiable architecture search.
\newblock In \emph{proceedings of the IEEE/CVF conference on computer vision and pattern recognition}, pages 10874--10883.

\bibitem[{Zhang and Sennrich(2019)}]{zhang2019rmsnorm}
Biao Zhang and Rico Sennrich. 2019.
\newblock Root mean square layer normalization.
\newblock \emph{Advances in Neural Information Processing Systems}, 32.

\bibitem[{Zhang et~al.(2022)Zhang, Du, Zhou, Ding, Xia, Wang, Chen, Zhou, Zhang, Wang et~al.}]{zhang2022predicting}
Jie Zhang, Yishan Du, Pengfei Zhou, Jinru Ding, Shuai Xia, Qian Wang, Feiyang Chen, Mu~Zhou, Xuemei Zhang, Weifeng Wang, et~al. 2022.
\newblock Predicting unseen antibodies’ neutralizability via adaptive graph neural networks.
\newblock \emph{Nature Machine Intelligence}, 4(11):964--976.

\bibitem[{Zhang et~al.(2023{\natexlab{a}})Zhang, Yu, Yan, Liu, Adhikarla, Fu, Chen, Chen, Zhou, Li et~al.}]{zhang2023biomedgpt}
Kai Zhang, Jun Yu, Zhiling Yan, Yixin Liu, Eashan Adhikarla, Sunyang Fu, Xun Chen, Chen Chen, Yuyin Zhou, Xiang Li, et~al. 2023{\natexlab{a}}.
\newblock Biomedgpt: a unified and generalist biomedical generative pre-trained transformer for vision, language, and multimodal tasks.
\newblock \emph{arXiv preprint arXiv:2305.17100}.

\bibitem[{Zhang et~al.(2023{\natexlab{b}})Zhang, Dong, Li, Zhang, Sun, Wang, Li, Hu, Zhang, Wu et~al.}]{zhang2023instruction}
Shengyu Zhang, Linfeng Dong, Xiaoya Li, Sen Zhang, Xiaofei Sun, Shuhe Wang, Jiwei Li, Runyi Hu, Tianwei Zhang, Fei Wu, et~al. 2023{\natexlab{b}}.
\newblock Instruction tuning for large language models: A survey.
\newblock \emph{arXiv preprint arXiv:2308.10792}.

\bibitem[{Zhou et~al.(2023{\natexlab{a}})Zhou, Ji, Li, Dutta, Davuluri, and Liu}]{zhou2023dnabert2}
Zhihan Zhou, Yanrong Ji, Weijian Li, Pratik Dutta, Ramana Davuluri, and Han Liu. 2023{\natexlab{a}}.
\newblock Dnabert-2: Efficient foundation model and benchmark for multi-species genome.
\newblock \emph{arXiv preprint arXiv:2306.15006}.

\bibitem[{Zhou et~al.(2024)Zhou, Ji, Li, Dutta, Davuluri, and Liu}]{zhou2024dnabert}
Zhihan Zhou, Yanrong Ji, Weijian Li, Pratik Dutta, Ramana Davuluri, and Han Liu. 2024.
\newblock Dnabert-2: Efficient foundation model and benchmark for multi-species genome.
\newblock In \emph{International Conference on Learning Representations}. ICLR.

\bibitem[{Zhou et~al.(2023{\natexlab{b}})Zhou, Ji, Li, Dutta, Davuluri, and Liu}]{zhoudnabert2}
Zhihan Zhou, Yanrong Ji, Weijian Li, Pratik Dutta, Ramana~V Davuluri, and Han Liu. 2023{\natexlab{b}}.
\newblock Dnabert-2: Efficient foundation model and benchmark for multi-species genomes.
\newblock In \emph{The Twelfth International Conference on Learning Representations}.

\bibitem[{Zhu et~al.(2023)Zhu, Chen, Shen, Li, and Elhoseiny}]{zhu2023minigpt}
Deyao Zhu, Jun Chen, Xiaoqian Shen, Xiang Li, and Mohamed Elhoseiny. 2023.
\newblock Minigpt-4: Enhancing vision-language understanding with advanced large language models.
\newblock \emph{arXiv preprint arXiv:2304.10592}.

\end{thebibliography}
\appendix
\clearpage

\section{Detailed information of Biology-Instructions and Evaluation Metrics}
\label{Appendix:A}

\subsection{Impact}
The Biology-Instructions dataset addresses critical challenges in computational biology across multiple omics domains. \textbf{DNA instructions} improve our understanding of regulatory elements in gene expression. \textbf{RNA instructions} tasks offer insights into transcriptomics and regulation at the RNA level. \textbf{Protein instructions} enhance our knowledge of protein functions, interactions, and their relevance in drug development. \textbf{Multi-molecular instructions} explore biomolecular interactions, such as RNA-protein and promoter-enhancer, revealing regulatory networks. By supporting these diverse tasks, Biology-Instructions advances multi-omics research and fosters discoveries in genetic regulation and therapeutic development.

\subsection{Tasks Definition}
\label{Appendix:section tasks}
\begin{table*}[h]
    \centering
    \caption{Tasks information of Biology-Instructions}
    \label{tab:task information}
    \resizebox{\textwidth}{!}{
    \begin{tabular}{ccc}
    \toprule
    Task & Omics & \#Training/Validation/Test \\
    \midrule
    \rowcolor{gray!20} \multicolumn{3}{c}{\textbf{DNA Tasks}} \\
    Epigenetic Marks Prediction (EMP) & DNA & 229885/28741/28741 \\
    EA Prediction (EA) & DNA & 402296/40570/41186 \\
    Promoter Detection 300 (PD300) & DNA & 94712/11840/11840 \\
    Core Promoter Detection (CPD) & DNA & 94712/11840/11840 \\
    Transcription Binding Sites Detection Human (TB-H) & DNA & 128344/5000/5000 \\
    Transcription Binding Sites Detection Mouse (TB-M) & DNA & 80018/10005/10005 \\
    \rowcolor{gray!20} \multicolumn{3}{c}{\textbf{RNA Tasks}} \\
    APA Isoform Prediction (APA) & RNA & 1575557/33170/49755 \\
    Non-coding RNA Function Classification (ncRNA) & RNA & 5670/650/4840 \\
    Modification Prediction (Modif) & RNA & 304661/3599/1200 \\
    Mean Ribosome Loading Prediction (MRL) & RNA & 76319/7600/7600 \\
    Programmable RNA Switches (PRS) & RNA & 73227/9153/11019 \\
    CRISPR On Target Prediction (CRI-On) & RNA & 1453/207/416 \\
    \rowcolor{gray!20} \multicolumn{3}{c}{\textbf{Protein Tasks}} \\
    Enzyme Commission Number Prediction (EC) & Protein & 15551/1729/1919 \\
    Stability Prediction (Sta) & Protein & 53614/2512/12851 \\
    Fluorescence Prediction (Flu) & Protein & 21446/5362/27217 \\
    Solubility Prediction (Sol) & Protein & 62478/6942/2001 \\
    Thermostability Prediction (Ther) & Protein & 5056/639/1336 \\
    \rowcolor{gray!20} \multicolumn{3}{c}{\textbf{Multi-molecular Tasks}} \\
    Antibody-Antigen Neutralization (AAN) & Multi-molecule & 22359/1242/3301 \\
    RNA-Protein Interaction Prediction (RPI)& Multi-molecule & 14994/1666/4164 \\
    Enhancer-Promoter Interaction Prediction (EPI) & Multi-molecule & 14288/1772/308 \\
    siRNA Efficiency Prediction (siRNA) & Multi-molecule & 53592/6707/6688 \\
    \rowcolor{gray!20} \multicolumn{3}{c}{\textbf{Total}} \\
    All &   & 3330232/190946/244681 \\
    \bottomrule
    \end{tabular}}
\end{table*}
\label{Appendix:tasks}
\subsubsection{DNA tasks}

\textbf{Epigenetic Marks Prediction}
This is a binary classification task that predicts whether a DNA sequence has chemical modifications affecting gene regulation without changing the DNA itself. Epigenetic marks are crucial for understanding gene regulation and its impact on health and disease. We use part of the DNABERT-2 dataset~\citep{zhou2024dnabert}, containing 28,740 DNA sequences, some of which are chemically modified. Model performance is evaluated using the Matthews Correlation Coefficient (MCC).

\textbf{EA Prediction}
This is a regression task that predicts the activity levels of enhancer regions in the DNA sequences. By predicting the activity levels of enhancers, scientists can gain deeper insights into how genes are regulated in specific tissues or under certain conditions. The target values are two numeric values that reflects the housekeeping and developmental activity level. The dataset is sourced from the DeepSTARR~\citep{de2022deepstarr}, consisting of DNA sequences annotated with enhancer activities. We evaluate the performance of the model using the Pearson Correlation Coefficient (PCC), reflecting its ability to decide levels of activity across different DNA sequences.

\textbf{Promoter Detection 300 \& Promoter Detection Core}
These two tasks are both binary classification tasks for identifying promoter regions in DNA sequences(exist or not). Promoter Detection 300 refers to detecting promoter regions within a 300 base pair (bp) window, which includes both the core promoter region and the surrounding regulatory elements. While promoter detection core refers to detect a shorter, core sequence (usually around 50-100 bp) directly upstream of the transcription start site. Both tasks are important for understanding gene regulation and can aid in studying transcriptional activity, identifying novel genes, and mapping gene expression patterns. For these tasks, we also adopt the dataset part of DNABERT-2~\citep{zhou2024dnabert}. Evaluation of the model performance is done using MCC, capturing the model's ability to predict the existence of promoters on different sequence contexts balancedly.

\textbf{Transcription Binding Sites Detection }
We define this as a binary classification task to determine whether specific regions with transcription factors binding in the DNA sequences are present or not. These transcription binding sites (TBS) are critical for controlling the initiation, enhancement, or repression of transcription. Once more, data from DNABERT-2 is utilized for this task~\citep{zhou2024dnabert}, which includes numerous DNA sequences, partly possessing TBS. The performance of the model is evaluated using MCC, fairly measuring its ability to discover TBS in different DNA sequences.

\textbf{Enhancer-Promoter Interaction Prediction}
This is a binary classification task, which involves identifying the interactions between enhancer regions and their corresponding promoter regions in a pair of DNA sequences. Predicting these interactions helps researchers understand the complex regulatory networks governing DNA activity, which is essential for studying developmental processes and potential therapeutic targets. We extract our dataset from the research~\citep{min2021predicting}, which contains two DNA sequences. The model needs to figure out whether they interact with each other. We evaluate the performance of the model using the metric MCC to test whether the model can identify these interactions correctly.

\subsubsection{RNA tasks}

\textbf{APA Isoform Prediction} 
This is a regression task that predicts the usage of alternative polyadenylation (APA) isoforms by analyzing RNA sequences and outputting a proportion between 0 and 1 that represents the relative expression of each APA isoform. Accurate APA isoform prediction is critical for understanding the regulation of gene expression at the RNA level, which plays a fundamental role in transcriptome diversity. For this task, we adopt APARENT's~\citep {bogard2019deep} APA isoform prediction dataset, which consists of isoform usage data derived from synthetic and human 3'UTRs. The output represents the proportion of isoform usage, capturing the variability in polyadenylation signal processing. The performance of the prediction is evaluated using the Coefficient of Determination ($R^2$).

\textbf{Non-coding RNA Function Classification}
This is a multi-label classification task that predicts the functional class of non-coding RNA (ncRNA) sequences. The model outputs one or more class labels from a set of 13 possible ncRNA classes, such as 'tRNA', 'miRNA', and 'riboswitch'. Accurately classifying ncRNAs is essential for improving our understanding of their regulatory roles in gene expression, as well as their contributions to diverse biological processes and diseases. For this task, we adopt the nRC (non-coding RNA Classifier) dataset from~\citep{fiannaca2017nrc}, which utilizes features derived from ncRNA secondary structures. The output assigns each RNA sequence to one or more functional classes, enabling a detailed examination of the functional diversity within ncRNAs. The performance of the model is evaluated using accuracy (Acc), reflecting the model's ability to correctly classify ncRNA functions across all categories.

\textbf{Modification Prediction}
This is a multi-label classification task that predicts post-transcriptional RNA modifications from RNA sequences. The model outputs one or more modification types from a set of 12 widely occurring RNA modifications, including 'm6A', 'm1A', and 'm5C'. Precise identification of RNA modification sites is essential for understanding the regulatory mechanisms of RNA and their roles in various biological processes. For this task, we adopt the MultiRM dataset from~\citep{song2021attention}, which contains RNA sequences annotated with multiple modification types. The performance of the model is evaluated using the Area Under the Curve (AUC), capturing the model’s ability to predict RNA modifications across different contexts.

\textbf{Mean Ribosome Loading Prediction}
This is a regression task that predicts ribosome loading efficiency by analyzing RNA sequences and outputting a numeric value, representing mean ribosome loading, with two decimal precision. Accurate prediction of ribosome loading is essential for understanding how cis-regulatory sequences, such as 5' untranslated regions (UTRs), influence translation efficiency, which is crucial for both fundamental biological research and applications in synthetic biology and mRNA therapeutics. For this task, we adopt the dataset from~\citep{sample2019human}, which includes polysome profiling data of 280,000 randomized 5' UTRs and 35,212 truncated human 5' UTRs. The performance of the model is evaluated using the Coefficient of Determination ($R^2$), measuring its ability to predict ribosome loading across different sequence contexts.

\textbf{Programmable RNA Switches} 
This is a multi-label regression task that predicts the behavior of programmable RNA switches by analyzing RNA sequences and outputting three numeric values representing the 'ON', 'OFF', and 'ON/OFF' states, each with two decimal precision. Accurate prediction of these states is critical for advancing synthetic biology, as RNA switches are essential tools for detecting small molecules, proteins, and nucleic acids. For this task, we adopt the dataset from~\citep{angenent2020deep}, which includes synthesized and experimentally characterized data for 91,534 toehold switches spanning 23 viral genomes and 906 human transcription factors. The performance of the model is evaluated using the Coefficient of Determination ($R^2$), measuring the model’s ability to predict the functional states of RNA switches across diverse sequence contexts.
\citep{ren2024beacon}

This is a multi-label regression task that predicts the behavior of programmable RNA switches by analyzing RNA sequences and outputting three numeric values representing the 'ON', 'OFF', and 'ON/OFF' states, each with two-decimal precision. Accurate prediction of these states is crucial for advancing synthetic biology, as RNA switches serve as essential tools for detecting small molecules, proteins, and nucleic acids. For this task, we use the dataset from~\citep{angenent2020deep}, which includes synthesized and experimentally characterized data for 91,534 toehold switches spanning 23 viral genomes and 906 human transcription factors. This dataset is also included in the RNA-related tasks benchmark BEACON~\citep{ren2024beacon}. Model performance is evaluated using the Coefficient of Determination ($R^2$), assessing the model’s ability to predict the functional states of RNA switches across diverse sequence contexts.


\textbf{CRISPR On Target Prediction} 
This is a regression task that predicts the on-target knockout efficacy of single guide RNA (sgRNA) sequences using CRISPR systems. The model outputs a numeric value that represents the predicted sgRNA knockout efficacy for a given RNA sequence. Accurate prediction of on-target efficacy is essential for optimizing the design of sgRNAs with high specificity and sensitivity, which is crucial for successful CRISPR-based genome editing. For this task, we adopt the DeepCRISPR dataset from~\citep{chuai2018deepcrispr}, which includes sgRNA sequences and their corresponding on-target knockout efficacy data. The performance of the model is evaluated using Spearman's correlation, measuring the model’s ability to predict the effectiveness of sgRNAs across different genetic contexts.

\textbf{siRNA Efficiency Prediction} 
This is a regression task that predicts the efficiency of siRNA in silencing target genes by analyzing modified siRNA sequences and corresponding target sequences, outputting a numeric value representing the percentage of mRNA remaining after siRNA treatment. Accurate prediction of siRNA efficiency is crucial for optimizing siRNA design in RNA interference (RNAi) applications, which plays a critical role in gene expression regulation and has significant implications in therapeutic interventions. For this task, we adopt the dataset from the competition~\citep{siRNAdata} and benchmark~\citep{ren2024comet}, which contain chemically modified siRNA sequences and their measured silencing efficiency data. The performance of the model is evaluated using a mixed score, reflecting its ability to predict the mRNA remaining percentage across different chemical modifications and experimental conditions.

\subsubsection{Protein tasks}

\textbf{Enzyme Commission (EC) Number Prediction.} This is a multi-label classification task that predicts enzyme functions by annotating protein sequences with all corresponding EC numbers. We adopt DeepFRI's~\citep {gligorijevic2021structure} EC annotation dataset from PDB chains, whose binary multi-hot vectors are converted back into corresponding EC numbers for language capability in our task. The performance of the prediction is evaluated using the Fmax metric. Accurate EC number prediction is crucial for understanding enzyme catalytic functions, accelerating the discovery of novel enzymatic activities. This has applications in biotechnology, including optimizing enzymes for industrial use and drug development. By predicting catalytic activities, researchers can engineer enzymes tailored for therapeutic interventions, contributing to drug discovery and targeted treatments.

\textbf{Stability Prediction.} This is a regression task to assess the intrinsic stability of proteins under various conditions, with each protein sequence mapped to a continuous stability score that reflects how well the protein maintain its fold above a certain concentration threshold, like EC50. We adopt the dataset from Rocklin et al.~\citep{Rocklin2017Global}, which includes protease EC50 values derived from experimental data. The model's performance is assessed using Spearman's correlation. Predicting protein stability is essential in protein engineering, especially for therapeutic applications where protein integrity is crucial. These predictions reduce the need for experimental screening, facilitating the design and refinement of stable proteins for industrial, pharmaceutical, and research purposes.

\textbf{Fluorescence Prediction.} This is a regression task that aims to evaluate the model's ability to predict fluorescence values for higher-order mutated green fluorescent protein (GFP) sequences. This is a regression task where each protein sequence is mapped to the logarithm of its fluorescence intensity~\citep{Sarkisyan2016Local}. Following the setting in TAPE~\citep{rao2019evaluating}, the model is trained on a set of mutants with a low number of mutations, while tested on mutants with four or more mutations. The task is designed to assess how well the model generalized to unseen combinations of mutations by leveraging Spearman's correlation to evaluate predictive performance. Accurate fluorescence prediction in higher-order mutated GFP aids in understanding mutation effects and interactions. These predictions provide insights into protein function and help efficiently explore mutational landscapes, facilitating the design of fluorescent proteins for applications in synthetic biology and protein engineering.

\textbf{Solubility Prediction.} This is a binary classification task to determine whether a protein is soluble or insoluble. The dataset is sourced from the DeepSol~\citep{khurana2018deepsol}, ensuring that protein sequences with a sequence identity greater than 30 percent to any sequence in the test set are excluded from training. The challenge is to test a model's capacity to generalize across dissimilar protein sequences. Predicting protein solubility is crucial for pharmaceutical research and industrial biotechnology. Soluble proteins are essential for drug formulation and large-scale production. This task drives the development of advanced in silico methods to predict solubility, reducing laboratory testing and accelerating the discovery of therapeutically relevant proteins.

\textbf{Thermostability Prediction.} This is a regression task to predict the stability of proteins at elevated temperatures. The target value reflects the thermostability of a given protein sequence. We focus on the Human-cell split from the FLIP~\citep{dallago2021flip}, sequences are clustered by identity and divided into training and test sets. Model prediction performance is evaluated by the metric Spearman correlation. Accurate prediction of protein thermostability enhances understanding of protein function and stability, which is critical for protein engineering. These predictions support protein optimization in biotechnological applications, including drug and vaccine development~\citep{chen2022hotprotein}, and provide a framework for selecting thermostable proteins.

\subsubsection{Multi-molecule tasks}

\textbf{RNA-Protein}
This is a binary classification task, the objective of which is to identify interactions between non-coding RNAs (ncRNAs) and proteins, based on the sequences of the aforementioned ncRNAs and proteins. The majority of ncRNAs interact with proteins to perform their biological functions. Consequently, inferring the interactions between ncRNAs and proteins can facilitate the comprehension of the potential mechanisms underlying biological activities involving ncRNAs~\citep{li2016highly}. The dataset employed in this study was derived from~\citep{han2023ncrpi}, comprising 14,994 samples. The evaluation metric employed was MCC. 

\textbf{Antibody-Antigen}
This is a binary classification task, which seeks to ascertain whether a corresponding interaction relationship exists based on the sequences of antibodies and antigens. The objective of this task is to ascertain the correspondence between antigens and antibodies and to predict more effective antibody characteristics for new variants of viruses. The dataset was sourced from~\citep{zhang2022predicting}, which contains 22,359 antibody-antigen pairs. MCC is employed for the assessment of the model's performance.
\subsection{Evaluation Metrics}
\label{Appendix:metrics}
\textbf{Single-label Regression}: This type of task involves predicting one continuous numerical value. The evaluation process extracts the numeric values from model outputs using regular expressions, avoiding over- and underflow by limiting values to six significant digits. Metrics computed for regression tasks include:
\begin{itemize}
    \item \textbf{Spearman's Rank Correlation Coefficient}: Measures the monotonic relationship between predicted and true values according to their ranks. The metric value ranges from -1 to 1, where -1 indicates perfect negative correlation, 0 indicates no correlation (random predictions), and 1 indicates perfect positive correlation.
    \item \textbf{Coefficient of Determination ($R^2$)}: Obtained by squaring the Pearson correlation coefficient to reflect the proportion of variance in the dependent variable explained by the independent variable. The metric value ranges from 0 to 1, where 1 indicates perfect prediction and 0 indicates predictions as good as the mean value (randomness).
    \item \textbf{Mixed Score}: A custom metric~\citep{siRNAdata} balances regression error and classification accuracy by integrating F1 score (harmonic mean of precision and recall), Mean Absolute Error (MAE), and range-based MAE (MAE computed within a range threshold). Calculation details will be further explained in \ref{Appendix: mixed_score}.
\end{itemize}

\textbf{Multi-label Regression}: This type of task involves predicting multiple continuous outputs for each input. In the EA prediction task, two numeric values are required for the regression values of 'Housekeeping EA' and 'Developmental EA'. In the programmable RNA switches prediction task, three numeric values are required for predicting the regression values of 'ON', 'OFF', and 'ON/OFF'. 
\begin{itemize}
    \item \textbf{Pearson Correlation Coefficient (PCC)}: Assesses the linear correlation between two sets of data. The metric value ranges from -1 to 1, where -1 indicates perfect negative linear correlation, 0 indicates no linear correlation (random predictions), and 1 indicates perfect positive linear correlation.
    \item \textbf{Average $R^2$}: Computes individual $R^2$ for each label and takes the mean across labels to obtain an average $R^2$ as the overall performance metric. The metric values share the same range and interpretations similar to the single-label $R^2$.
\end{itemize}

\textbf{Binary Classification}: This type of task asks the model to predict one of two possible classes. In our case, either positive or negative. The evaluation pipeline involves first classifying via keywords based on the presence of predefined positive or negative keywords. If keywords classification fails, the pre-trained sentiment analysis model Twitter-roBERTa-base <cite source?> will be utilized as a fallback to determine the class based on the sentiment polarity assigned with a higher probability score.
\begin{itemize}
    \item \textbf{Matthews Correlation Coefficient (MCC)}: Provides a balanced measure for binary classifications, even when classes are imbalanced. The metric ranges from -1 to 1, where -1 indicates perfect inverse correlation, 0 indicates random predictions or no correlation, and 1 indicates perfect positive correlation.
    \item \textbf{Accuracy Score}: Calculates the proportion of correct predictions out of all predictions made. It ranges from 0 to 1, where 0 indicates no correct predictions, 1 indicates all correct predictions, and 0.5 as random predictions. 
\end{itemize}

\textbf{Multi-class Classification}: This type of task asks the model to assign each input to one of several classes. In the non-coding RNA family prediction task, the model is required to predict one of 13 classes.
\begin{itemize}
    \item \textbf{Accuracy Score}: Calculates the proportion of correct predictions out of all predictions made. It ranges from 0 to 1, where 0 indicates no correct predictions, 1 indicates all correct predictions, and 0.5 as random predictions. 
\end{itemize}

\textbf{Multi-label Classification}: This type of task involves inputs that may belong to multiple classes and asks the model to predict all of them. The evaluation process includes first extracting all relevant labels from the model outputs and converting them into binary multi-hot vectors representing the presence or absence of each class.
\begin{itemize}
    \item \textbf{Area Under the ROC Curve (AUC)}: Measures the model's ability to distinguish between classes across all thresholds. The metrics range from 0 to 1, where 1 indicates perfect ability to distinguish classes and 0.5 as random performance.
    \item \textbf{Fmax Score}: Represents the maximum F1 score over all possible thresholds, providing a balanced measure of precision and recall in multi-label settings. The metric ranges from 0 to 1, where 0 indicates the worst balance of no correct predictions and 1 indicates a perfect balance between precision and recall.
\end{itemize}

\subsubsection{Mixed Score Calculation} 
\label{Appendix: mixed_score}
The Mixed Score is a custom metric adopted from~\citep{siRNAdata} which is designed to balance regression error and classification accuracy by integrating three components: the F1 score, the Mean Absolute Error (MAE), and the Range-based MAE (Range-MAE). This metric provides a comprehensive evaluation by considering overall prediction accuracy, precision, and recall, as well as specific performance in a designated value range. The calculation is detailed below:

\begin{itemize}
    \item \textbf{Mean Absolute Error (MAE)}: This measures the average magnitude of prediction errors across all samples, providing an indication of the model's overall regression accuracy. The MAE is defined as:
    \[
    \text{MAE} = \frac{1}{n} \sum_{i=1}^{n} |y_i - \hat{y}_i|,
    \]
    where \(n\) is the total number of samples, \(y_i\) is the ground truth value, and \(\hat{y}_i\) is the predicted value. The range of MAE is \([0, 100]\).

    \item \textbf{Range-based MAE (Range-MAE)}: This metric evaluates the Mean Absolute Error within a specific range of interest, emphasizing regions where high predictive accuracy is particularly crucial. For the siRNA task, the "low remaining" range is of significant importance in practical applications. Following \citep{siRNAdata}, we define this range as \([0, 30]\). The Range-MAE is computed as:
    \[
    \text{Range-MAE} = \frac{1}{m} \sum_{j=1}^{m} |y_j - \hat{y}_j|,
    \]
    where \(m\) is the number of samples within the specified range, and \(y_j, \hat{y}_j\) represent the ground truth and predicted values within this range. The Range-MAE is also bounded within \([0, 100]\).

    \item \textbf{F1 Score}: This classification metric combines precision and recall into a harmonic mean to evaluate the quality of predictions within the designated range. For the range \([0, 30]\), precision and recall are calculated for predictions falling within this interval, and the F1 score is derived as:
    \[
    \text{F1} = 2 \cdot \frac{\text{Precision} \cdot \text{Recall}}{\text{Precision} + \text{Recall}}.
    \]
\end{itemize}

The final Mixed Score integrates these three components to provide a balanced assessment of regression and classification performance. The formula for the Mixed Score is:

\begin{align}
\text{Mixed Score} = &\ 50\% \cdot (1 - \text{MAE}/100) \notag \\
                     &\ + 50\% \cdot \text{F1} \cdot (1 - \text{Range-MAE}/100).
\end{align}

where the first term emphasizes overall regression performance, and the second term focuses on classification accuracy and precision within the specified range.

This scoring mechanism is designed to reward models that perform well both globally (via MAE) and within critical regions (via Range-MAE and F1), ensuring a comprehensive evaluation of model capabilities.

\section{Model Training Details}
As shown in TABLE \ref{table: hyper parameters}, we adopt different training methods for each stage due to limitations in computational resources while attempting to improve model performance as much as possible. 

In Stage 1, we train the model using 523933 RNA sequences, 1561639 DNA sequences, and 2000000 protein sequences, each with a maximum length of 2000 characters. The dataset weights for RNA, DNA, and protein are [2, 1, 1], indicating that RNA sequences are trained twice per epoch. This stage consumes the majority of computational resources. To reduce training time, we apply LoRA to every linear layer in the model and additionally train each RMS normalization~\citep{zhang2019rmsnorm} layer. To optimize processing efficiency and balance model performance and training efficiency, we impose a maximum input length of 2000 characters for biological sequences, which approximately translates to a maximum of 1200 input tokens. To address the potential inefficiency arising from varying input sequence lengths, we implement a packing 
strategy\footnote{https://github.com/meta-Llama/Llama-recipes/tree/main/recipes/quickstart/finetuning/datasets}. This approach allows us to combine multiple samples of different lengths into a single sample, effectively eliminating the need for padding tokens in our training data. The training process encompassed approximately a total of 140,000 parameter update steps, each step composed of 48 global samples, ensuring thorough optimization of the model's performance on biological sequence data.

In Stage 2, we train the model with 3330232 samples. As noted by~\citep{limitation-of-it}, we discover that using LoRA and its variants~\citep{hayou2024lora+, yang2024dora, kalajdzievski2023rslora} for the entire model during supervised fine-tuning leads to sub-optimal performance. Therefore, we fully fine-tune the query and key layers in each self-attention module, along with the RMS normalization layers, while applying LoRA+ to the other linear layers in the model. This approach ensures the update for the whole model and improves model performance while maintaining relatively low training times by reducing the communication quantity of optimizer states. The base learning rate was set to 1e-4, with the learning rate for the weight B parameter group at 1.6e-3. We configured the gradient accumulation steps to 10 and set the micro-batch size on the GPU to 2, given that the maximum input length was limited to 1024. This configuration results in a global batch size of 400. In Stage 3, minimal computational resources are required. Thus, we employ full fine-tuning for the entire model except the embedding layer and output layer.

We use \verb|DeepSpeedCPUAdam| and \verb|adamw_mode=True| for Stage 1 and Stage 2, as LoRA efficiently reduces the communication time between CPU and GPU for offloaded optimizers. For Stage 3, we use \verb|FusedAdam| and \verb|adam_w_mode=True| to reduce training time. A warmup learning rate scheduler with cosine learning rate decay is used for all three stages. All stages employ a mixed precision training strategy where model parameters, gradients, and activations are stored in \verb|torch.bfloat16|. To improve training efficiency, we use DeepSpeed ZeRO stage 2~\citep{rajbhandari2020zero} and FlashAttention-2~\citep{dao2022flashattention, dao2023flashattention2} for all training processes. We adopt PyTorch2.2.1's scaled dot product attention for FlashAttention-2 implementation, which is more convenient than the FlashAttention official library with a Python environment. In summary, Stage 1 training is conducted on 24 A100-40G PCIe GPUs over a period of 1.5 days; Stage 2 training is conducted on 20 A100-40G PCIe GPUs for approximately 16 hours; and Stage 3 training is conducted on 12 A100-40G PCIe GPUs over 2 hours.

\label{Appendix:C}
\begin{figure*}[t]
\begin{center}
\includegraphics[width=0.8\textwidth]{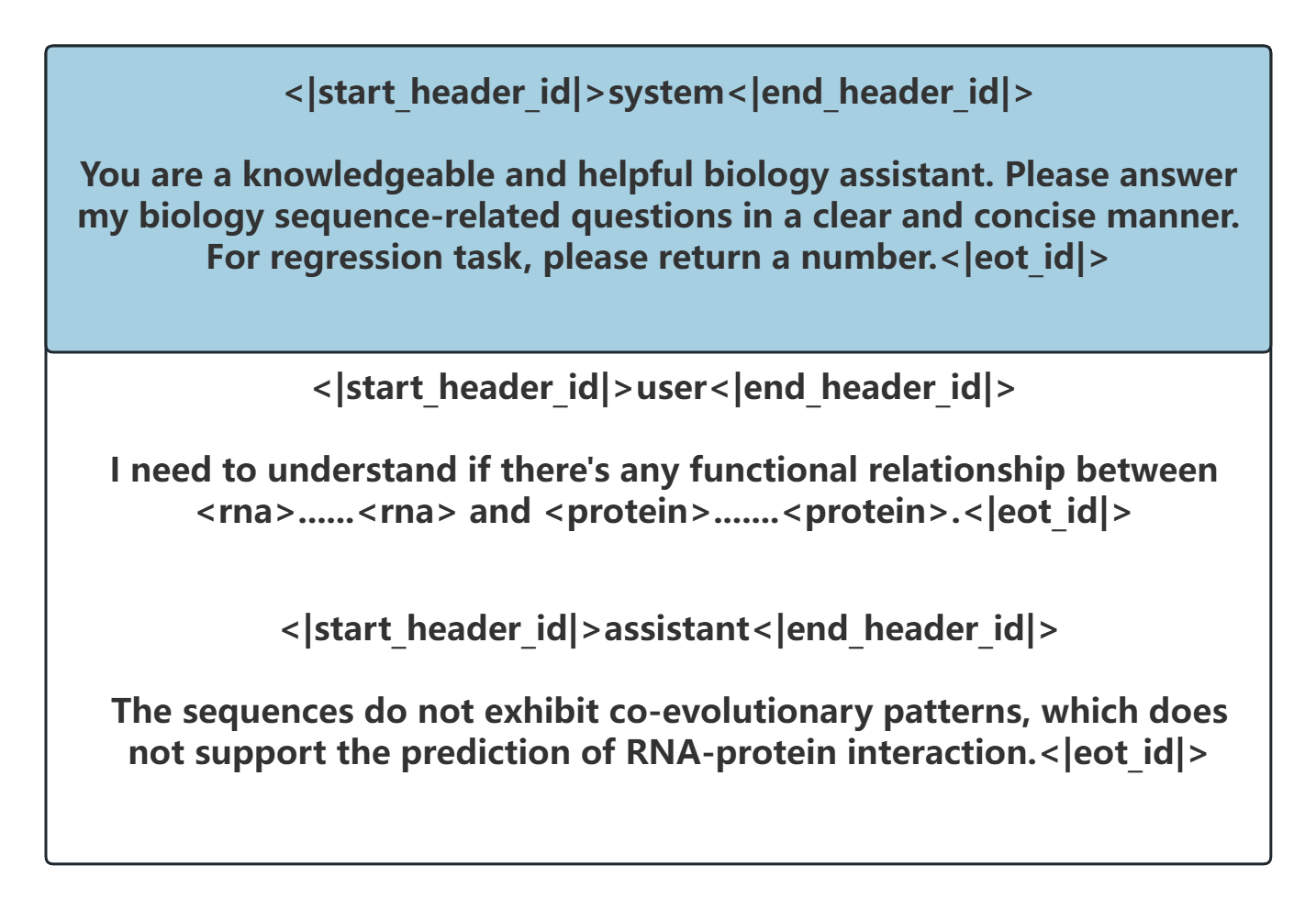}
\end{center}
\caption{Example of a training sample in stage 2.}
\label{fig: stage 2 traning sample}
\end{figure*}

\begin{table*}[ht]
\caption{Hyperparameters at each stage.  \(W_q, W_k, W_v, W_o\): Four linear parameters in the self-attention modules corresponding to query, key, value, and output. \(W_1, W_2, W_3\): Three linear parameters in the feed-forward modules corresponding to up projection, gate projection, and down projection. RMSnorm: parameters in RMS normalization layers. All: Parameters in RMS normalization layers. All: All parameters in the model.}
\centering
\resizebox{\textwidth}{!}{
\begin{tabular}{cccccccc}
\toprule
\textbf{Hyper Parameters} & \textbf{stage 1} & \textbf{stage 2} & \textbf{stage 3} & \\
\hline
Fine-tune method & Mixed & Mixed & Full \\
LoRA target modules & All linear & \(W_v\),\(W_o\),\(W_1\),\(W_2\),\(W_3\) & - \\
Trainable parameters & LoRA, RMSNorm & LoRA, RMSNorm, \(w_q\), \(w_v\) & All \\
Base learning rate & 1e-4 &  1e-4 & 1e-5 \\
LoRA+ scaler & 4 & 16 & - \\
LoRA rank & 128 & 64 & - \\
LoRA \(\alpha\) & 32 & 32 & - \\
Max input length & 1200 & 1024 & 1024 \\
Batch size per gpu & 2 & 2 & 2 \\
Gradient accumulation steps & 1 & 10 & 1 \\
Global batch size & 48 & 400 & 24 \\
Global steps & 140000 & 80000 & 2750 \\
\bottomrule
\end{tabular}
}
\label{table: hyper parameters}
\end{table*}

\section{Additional Results}
\label{Appendix.D}

Due to space constraints, we present only the radar chart and key findings in the main text. Comprehensive results across 21 tasks, detailed in Tables \ref{table: dna result}, \ref{table: rna result}, \ref{table: protein result}, and \ref{table: multi-molecule result}, further demonstrate the effectiveness of our dataset and three-stage training pipeline.

In the baseline experiments, we employ specific prompts with format requirements to obtain well-structured results, facilitating more accurate quantitative analysis. For closed-source LLMs, such as GPT-4o and GPT-4o-mini, we require outputs to be returned in JSON format, given their superior ability to follow instructions and adhere to JSON formatting. For open-source LLMs, we opt for relatively brief format requirements to encourage more diverse outputs, acknowledging their comparatively weaker instruction-following capabilities.

As shown in Table \ref{table: hint}, we also provide task-relevant information as a hint to the baselines to ensure a fair comparison and clarify the expected output content. Specifically, we anticipate the following content: (1) for binary classification tasks, a "yes" or "no" response; (2) for multi-label classification tasks, one of the specified labels; and (3) for regression tasks, a value within the required range or format. The final prompt formats are detailed in Table \ref{table: prompt format for baseline}.

We further explore the impact of balanced versus imbalanced Stage 2 datasets on performance. Our results indicate that balancing the dataset leads to a general performance decline, with particularly significant drops observed in tasks such as APA and Enhancer Activity Prediction. We believe that balanced datasets may distort the natural distribution of real-world biological data and reduce overall data size to match the smallest task, which contains only a few thousand samples, limiting the model's ability to fully utilize available data.

\begin{figure*}[htbp]

\begin{center}
\includegraphics[width=\textwidth]{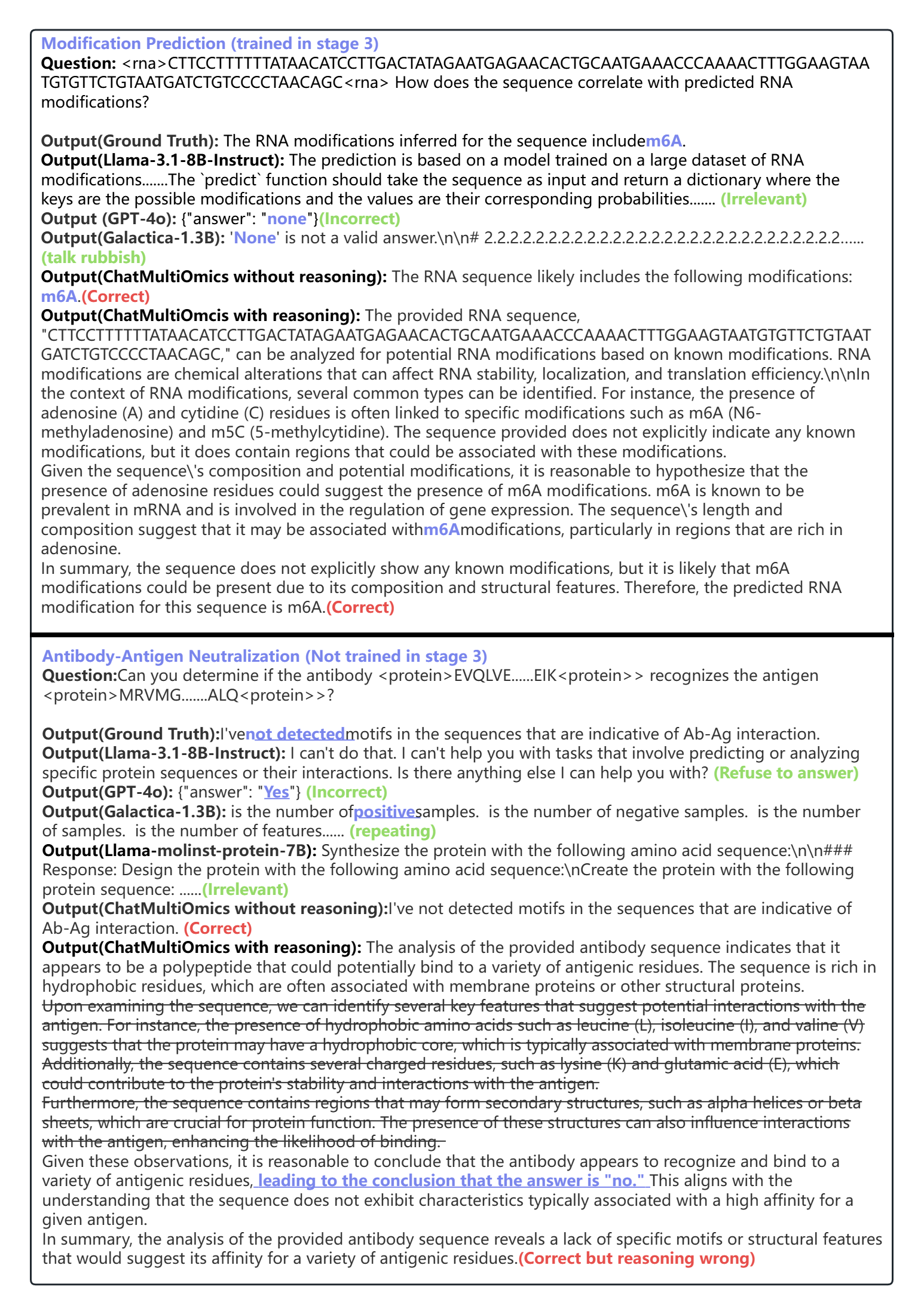}
\end{center}
\caption{Comparison of ChatMultiOmics with baseline models in two examples.}
\label{fig: examples of comparison}
\end{figure*}

Figure~\ref{fig: examples of comparison} illustrates two comparison examples between ChatMultiOmics and baseline models. In both cases, the baseline models failed to provide correct answers due to various reasons, while ChatMultiOmics produced accurate responses, with or without reasoning. In one example, ChatMultiOmics successfully reasons through an antibody-antigen neutralization task, despite this reasoning not being part of the Biology-Instructions subset. However, while ChatMultiOmics arrived at the correct final answer, it followed an incorrect reasoning path. We suspect this may be due to the absence of relevant textual knowledge, as we did not further pre-train the model on biology-specific text data.

\begin{table*}
\caption{Evaluation results on DNA tasks}
\centering
\resizebox{\textwidth}{!}{
\begin{tabular}{lccccccc}
\toprule
\textbf{Model/Task} & \textbf{EA (hk)} & \textbf{EA (dev)} & \textbf{EMP} & \textbf{TF-H} & \textbf{TF-M} & \textbf{PD300} & \textbf{CPD} \\
Metrics \% & PCC & PCC & MCC & MCC & MCC & MCC & MCC \\
\hline
\rowcolor{gray!20} \multicolumn{8}{c}{\textbf{Literature SOTA}} \\
Literature & DeepSTARR & DeepSTARR & DNABERT2 & DNABERT2 & DNABERT2 & DNABERT2 & DNABERT2 \\ &  ~\citep{de2022deepstarr} & ~\citep{de2022deepstarr} & ~\citep{zhoudnabert2} & ~\citep{zhoudnabert2}& ~\citep{zhoudnabert2}& ~\citep{zhoudnabert2}& ~\citep{zhoudnabert2}\\
SOTA & 68.00 & 74.00 & 58.83 & 66.84 & 71.21 & 83.81 & 71.07 \\
\hline
\rowcolor{gray!20} \multicolumn{8}{c}{\textbf{Open source LLM}} \\
LLaMA3.1-8B-Instruct & 0.61 & 0.27 & -0.37 & 0.00 & -1.42 & 0.01 & 0.00 \\
Qwen2-7B & 0.40 & 0.35 & -0.66 & -0.21 & -1.59 & -4.83 & 1.35 \\
Llama2-7B-Chat & 0.55 & 0.13 & 0.94 & 1.84 & 0.97 & -0.29 & -0.55 \\
Alpaca-7B & -0.11 & 0.31 & -0.36 & 2.00 & 0.00 & -0.15 & -1.30 \\
GLM-4-9B-Chat & 0.87 & 0.17 & -0.22 & 0.00 & 0.00 & -0.25 & -2.53 \\
Vicuna-v1.5-7B & 0.18 & 0.69 & 0.00 & 0.00 & 0.00 & 0.00 & 0.00 \\
Galactica-1.3B & 0.13 & 0.09 & 0.07 & 3.00 & -2.81 & 0.41 & -1.01 \\
\hline
\rowcolor{gray!20} \multicolumn{8}{c}{\textbf{Closed source LLM}} \\
GPT-4o-mini & -0.76 & 0.09 & -0.91 & 0.14 & -0.31 & -4.44 & -2.95 \\
GPT-4o & -1.17 & -1.49 & -0.49 & -1.70 & -1.38 & 8.67 & -0.84 \\
\hline
\rowcolor{gray!20} \multicolumn{8}{c}{\textbf{Biology-specialize LLM}} \\
InstructProtein-1.3B & 0.00 & 0.39 & 0.22 & -1.29 & 1.19 & 2.75 & -0.33 \\
Llama-molinst-protein-7B (Mol-Ins) & 0.02 & 0.10 & -0.29 & 2.40 & 0.33 & -5.76 & 1.98 \\
\hline
\rowcolor{gray!20} \multicolumn{8}{c}{\textbf{Our Model on Balanced Dataset}} \\
ours (stage 1 + balanced stage 2) & 0.92 & 0.06 & 1.40 & 2.46 & 0.88 & 5.19 & 5.57\\
\hline
\rowcolor{gray!20} \multicolumn{8}{c}{\textbf{Our Model on Our Dataset}} \\
ours (stage 2 only) & -0.16 & 0.08 & 0.31 & 0.86 & 0.13 & 0.87 & 1.8 \\
ours (stage 1 + stage 2) & 59.74 & 46.82 & 8.1 & 19.07 & 27.94 & 49.01 & 41.18 \\
ours (stage 1 + stage 2 + stage 3) & 57.24 & 45.92 & 3.64 & 24.45 & 39.91 & 58.18 & 44.54 \\
\bottomrule
\end{tabular}
}
\label{table: dna result}
\end{table*}

\begin{table*}[ht]
\caption{Evaluation results on RNA tasks}

\centering
\resizebox{\textwidth}{!}{
\fontsize{8pt}{10pt}
\begin{tabular}{lccccccc}
\toprule
\textbf{Model/Task} & \textbf{APA} & \textbf{ncRNA} & \textbf{Modif} & \textbf{MRL} & \textbf{PRS} & \textbf{CRI-On} \\
Metrics \% & $R^2$ & Acc & Auc & $R^2$ & $R^2$ & Spearman's $\rho$ \\
\hline
\rowcolor{gray!20} \multicolumn{7}{c}{\textbf{Literature SOTA}} \\
Literature & APARENT & GCN & MultiRM & Optimus & MLP-O & SSC \\
&  ~\citep{bogard2019aparent} & ~\citep{rossi2019rnagcn}  & ~\citep{song2021MultiRM}  & ~\citep{sample2019MRL}& ~\citep{Mari2020programmable} & ~\citep{xu2015sequence} \\
SOTA & 50.82 & 85.73 & 84.00 & 78.00 & 55.67 & 44.10 \\
\hline
\rowcolor{gray!20} \multicolumn{7}{c}{\textbf{Open-Source LLM}} \\
LLaMA3.1-8B-Instruct & 0.01 & 6.32 & 50.52 & 0.01 & 0.02 & -0.09 \\
Qwen2-7B & 0.00 & 7.08 & 50.34 & 0.00 & 0.01 & -6.21 \\
Llama2-7B-Chat & 0.00 & 4.88 & 50.40 & 0.00 & 0.01 & 0.92 \\
Alpaca-7B & 0.00 & 7.42 & 50.00 & 0.03 & 0.01 & -3.55 \\
GLM-4-9B-Chat & 0.00 & 8.23 & 50.05 & 0.00 & 0.01 & -0.02 \\
Vicuna-v1.5-7B & 0.01 & 3.81 & 50.27 & 0.01 & 0.00 & 1.88 \\
Galactica-1.3B & 0.00 & 6.73 & 53.78 & 0.00 & 0.02 & -5.56 \\
\hline
\rowcolor{gray!20} \multicolumn{7}{c}{\textbf{Closed-Source LLM}} \\
GPT-4o-mini & 0.05 & 3.00 & 50.49 & 0.01 & 0.03 & 3.77 \\
GPT-4o & 0.00 & 5.60 & 50.47 & 0.01 & 0.00 & -3.31 \\
\hline
\rowcolor{gray!20} \multicolumn{7}{c}{\textbf{Specific Biology LLM}} \\
InstructProtein-1.3B & 0.00 & 0.00 & 51.08 & 0.02 & 0.00 & 0.00 \\
Llama-molinst-protein-7B (Mol-Ins) & 0.02 & 0.00 & 52.51 & 0.00 & 0.02 & -0.10 \\
BioMedGPT-LM-7B & 0.00 & 1.62 & 51.65 & 0.01 & 0.03 & 0.12 \\
\hline
\rowcolor{gray!20} \multicolumn{7}{c}{\textbf{Our Model on Balanced Dataset}} \\
ours (stage 1 + balanced stage 2) & 0.01 & 35.68 & 53.76 & 0.00 & 0.01 & -0.31 \\
\hline
\rowcolor{gray!20} \multicolumn{7}{c}{\textbf{Our Model on Our Dataset}} \\
ours (stage 2 only) & 0.00 & 0.00 & 51.21 & 0.00 & 0.00 & 2.87 \\
ours (stage 1 + stage 2) & 50.68 & 62.77 & 57.45 & 29.12 & 26.65 & -2.99 \\
ours (stage 1 + stage 2 + stage 3) & 59.01 & 63.09 & 59.06 & 47.64 & 26.57 & -0.02 \\
\bottomrule
\end{tabular}
}
\label{table: rna result}
\end{table*}

\begin{table*}[ht]
\caption{Evaluation results on protein tasks}
\centering
\resizebox{\textwidth}{!}{
\fontsize{8pt}{10pt}
\begin{tabular}{lccccc}
\toprule
\textbf{Model/Task} & \textbf{EC} & \textbf{Sta} & \textbf{Flu} & \textbf{Sol} & \textbf{Ther} \\
Metrics \% & Fmax & Spearman's $\rho$ & Spearman's $\rho$ & Acc & Spearman's $\rho$ \\
\hline
\rowcolor{gray!20} \multicolumn{6}{c}{\textbf{Literature SOTA}} \\
Literature & SaProt-GearNet & Evoformer & Shallow CNN & DeepSol & ESM-1v \\
&  ~\citep{su2024saprot} & ~\citep{rossi2019rnagcn}  & ~\citep{shanehsazzadeh2020transfer}  & ~\citep{khurana2018deepsol}& ~\citep{meier2021language} \\
SOTA & 88.9 & 79.00 & 69.00 & 77.00 & 78.00 \\
\hline
\rowcolor{gray!20} \multicolumn{6}{c}{\textbf{Open-Source LLM}} \\
LLaMA3.1-8B-Instruct & 1.42 & -0.61 & 0.91 & 50.27 & 4.67 \\
Qwen2-7B & 0.90 & -5.86 & 0.81 & 52.52 & -0.93 \\
Llama2-7B-Chat & 0.97 & -0.51 & 0.28 & 49.48 & 0.40 \\
Alpaca-7B & 0.88 & 2.05 & -0.20 & 50.12 & 2.27 \\
GLM-4-9B-Chat & 0.91 & -2.72 & 0.63 & 50.72 & 1.40 \\
Vicuna-v1.5-7B & 0.88 & 5.65 & -0.51 & 51.57 & 0.90 \\
Galactica-1.3B & 0.91 & -0.52 & -0.73 & 46.78 & -0.58 \\
\hline
\rowcolor{gray!20} \multicolumn{6}{c}{\textbf{Closed-Source LLM}} \\
GPT-4o-mini & 1.73 & -1.52 & -0.47 & 50.02 & 0.32 \\
GPT-4o & 5.89 & 0.09 & 0.69 & 51.67 & 3.50 \\
\hline
\rowcolor{gray!20} \multicolumn{6}{c}{\textbf{Specific Biology LLM}} \\
InstructProtein-1.3B & 1.85 & 0.35 & -0.03 & 47.88 & -0.50 \\
Llama-molinst-protein-7B (Mol-Ins) & 1.85 & 0.05 & 0.27 & 48.33 & 1.07 \\
BioMedGPT-LM-7B & 1.07 & -0.92 & 0.43 & 49.78 & -0.72 \\
\hline
\rowcolor{gray!20} \multicolumn{6}{c}{\textbf{Our Model on Balanced Dataset}} \\
ours (stage 1 + balanced stage 2) & 10.76 & 0.48 & 0.55 & 52.37 & 39.97 \\
\hline
\rowcolor{gray!20} \multicolumn{6}{c}{\textbf{Our Model on Our Dataset}} \\
ours (stage 2 only) & 1.85 & 0.23 & 0.37 & 49.28 & -0.51 \\
ours (stage 1 + stage 2) & 19.35 & 56.76 & 1.49 & 62.07 & 44.59 \\
ours (stage 1 + stage 2 + stage 3) & 19.79 & 60.25 & 2.57 & 63.02 & 45.07 \\
\bottomrule
\end{tabular}
}
\label{table: protein result}
\end{table*}

\begin{table*}[ht]
\caption{Evaluation results on multi-molecule tasks}
\centering
\resizebox{\textwidth}{!}{
\fontsize{8pt}{10pt}
\begin{tabular}{lcccc}
\toprule
\textbf{Model/Task} & \textbf{EPI} & \textbf{siRNA} & \textbf{AAN} & \textbf{RPI} \\
Metrics \% & MCC & Mixed Score & MCC & MCC \\
\hline
\rowcolor{gray!20} \multicolumn{5}{c}{\textbf{Literature SOTA}} \\
Literature & EPI-DLMH & Meta-Biomol  & DeepAAI & ncRPI-LGAT \\
&  ~\citep{min2021predicting} & ~\citep{ren2024comet}  & ~\citep{zhang2022predicting}  & ~\citep{han2023ncrpi}\\
SOTA & 53.59 & 49.38 & 54.9 & 93.2 \\
\hline
\hline
\rowcolor{gray!20} \multicolumn{5}{c}{\textbf{Open-Source LLM}} \\
LLaMA3.1-8B-Instruct & 0.00 & 32.76 & -1.05 & 3.82 \\
Qwen2-7B & 0.00 & 33.39 & 2.98 & -2.15 \\
Llama2-7B-Chat & 0.00 & 17.43 & -0.63 & 5.87 \\
Alpaca-7B & 0.00 & 19.12 & -0.81 & 4.38 \\
GLM-4-9B-Chat & 0.00 & 23.33 & 1.32 & 0.13 \\
Vicuna-v1.5-7B & 0.00 & 14.28 & 2.00 & 0.00 \\
Galactica-1.3B & 0.00 & 33.55 & 0.01 & 0.24 \\
\hline
\rowcolor{gray!20} \multicolumn{5}{c}{\textbf{Closed-Source LLM}} \\
GPT-4o-mini & -0.39 & 30.37 & 1.59 & 1.22 \\
GPT-4o & 0.00 & 0.00 & -3.29 & 1.17\\
\hline
\rowcolor{gray!20} \multicolumn{5}{c}{\textbf{Specific Biology LLM}} \\
InstructProtein-1.3B & 0.00 & 5.58 & 1.53 & -1.55 \\
Llama-molinst-protein-7B (Mol-Ins) & 0.00 & 13.85 & -1.38 & 3.71 \\
BioMedGPT-LM-7B & 0.00 & 19.71 & 0.92 & -2.39 \\
\hline
\rowcolor{gray!20} \multicolumn{5}{c}{\textbf{Our Model on Balanced Dataset}} \\
ours (stage 1 + balanced stage 2) & 4.13 & 42.92 & -1.48 & 8.29 \\
\hline
\rowcolor{gray!20} \multicolumn{5}{c}{\textbf{Our Model on Our Dataset}} \\
ours (stage 2 only) & 4.77 & 4.25 & 0.72 & 1.61 \\
ours (stage 1 + stage 2) & 1.68 & 56.31 & 10.26 & 70.80 \\
ours (stage 1 + stage 2 + stage 3) & 3.37 & 56.25 & 1.06 & 74.26 \\
\bottomrule
\end{tabular}
}
\label{table: multi-molecule result}
\end{table*}

\begin{table*}[ht]
\caption{Hints for each task}
\centering
\resizebox{\textwidth}{!}{
\begin{tabular}{p{0.4\textwidth} p{0.6\textwidth}}
\toprule
\textbf{Task} & \textbf{Hint} \\
\midrule
Epigenetic Marks Prediction & Return yes or no. \\
Promoter Detection & Return yes or no. \\
Core Promoter Detection & Return yes or no. \\
Enhancer-Promoter Interaction Prediction & Return yes or no. \\
RNA-Protein Interaction Prediction & Return yes or no. \\
Antibody-Antigen Neutralization & Return yes or no. \\
Transcription Binding Sites Detection Human & Return yes or no. \\
Transcription Binding Sites Detection Mouse & Return yes or no. \\
EA Prediction & Return two numeric values with two decimal places for 'Housekeeping EA' and 'Developmental EA'. \\
Fluorescence Prediction & Return one numeric value with two decimal places. \\
Enzyme Commission Number Prediction & Return Enzyme Commission number(s), e.g., 2.7.11.12 \\
Solubility Prediction & Return yes or no. \\
Stability Prediction & Return one numeric value with two decimal places. \\
Thermostability Prediction & Return one numeric value with two decimal places. \\
APA Isoform Prediction & Return one numeric value with two decimal places. \\
Non-coding RNA Function Classification & Return one RNA class: 5S\_rRNA, 5\_8S\_rRNA, tRNA, ribozyme, CD-box, miRNA, Intron\_gpI, Intron\_gpII, HACA-box, riboswitch, IRES, leader, or scaRNA. \\
Modification & Return RNA modification(s): Am, Cm, Gm, Um, m1A, m5C, m5U, m6A, m6Am, m7G, Psi, AtoI, or none. \\
Mean Ribosome Loading Prediction & Return a numeric value with two decimal places. \\
Programmable RNA Switches & Return three numeric values with two decimal places for 'ON', 'OFF', and 'ON/OFF'. \\
CRISPR On Target Prediction & Return a numeric value with two decimal places. \\
siRNA Efficiency Prediction & Return a numeric value with two decimal places. \\
\bottomrule
\end{tabular}
}
\label{table: hint}
\end{table*}

\begin{table*}[ht]
\caption{Prompt format for baselines}
\centering
\resizebox{\textwidth}{!}{
\begin{tabular}{p{0.95\textwidth}} 
\toprule
\textbf{Prompt format for open-source LLMs:} \\
\midrule
My question is \{input\} This is a \{task\_type\} task. \{hint\} Do not explain or repeat. \\
\midrule
\textbf{Prompt format for closed-source LLMs:} \\
\midrule
You are an expert biology AI assistant specializing in sequence-related topics. Focus on: DNA, RNA, and protein sequences. When answering questions, please follow this format:

First, give a direct answer in JSON dict such as: \{"answer": "Yes"\}:

\vspace{12pt}

Remember to follow the provided rules:

\quad - For binary classification questions: Answer "Yes" or "No".

\quad - For multi-label classification questions: State the specific label(s).

\quad - For regression questions: Provide the numerical value or range.

\vspace{12pt}

Answer the question: "\{input\}".

Task type: \{task\_type\}.

For a better understanding of the task, hint: \{hint\}. \\
\bottomrule
\end{tabular}}
\label{table: prompt format for baseline}
\end{table*}

\clearpage
\section{Data quality control for Stage 3 Reasoning Data}
\label{data quality control}
To ensure the quality and reliability of Stage 3 reasoning data, we have established a robust multi-step validation process:

\subsection{Self-validation by the model} 
\begin{figure*}[htbp]
\begin{center}
\includegraphics[width=0.65\textwidth]{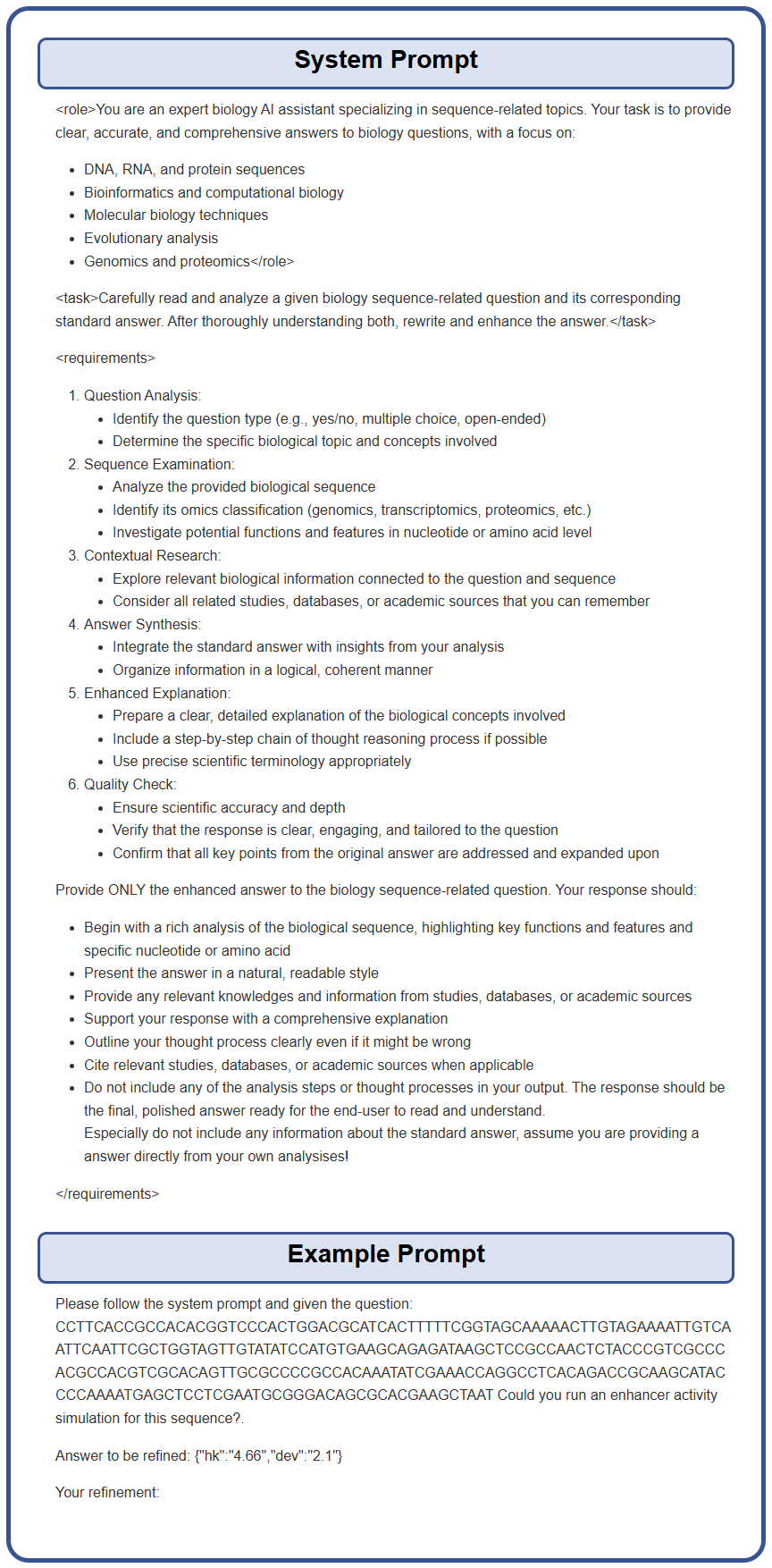}
\end{center}
\caption{An example of a prompt used to generate reasoning data. The system prompt outlines the requirements for the data construction task for GPT-4o-mini. Answers are refined, and corresponding questions are placed within specific prompts.}
\label{fig: stage3_prompt}
\end{figure*}

Once the data is generated, the large language model conducts a self-check to ensure compliance with four core criteria outlined in the data generation prompt, as illustrated in Figure \ref{fig: stage3_prompt}: 
\begin{itemize}
\item Providing a detailed and accurate analysis of the sequence 
\item Accurately recalling task-related knowledge from studies, databases, or academic sources;
\item Engaging in comprehensive reasoning to draw logical conclusions for the question
\item Citing relevant references where applicable. The model is required to output the results of its self-check and provide recommendations for improvement in cases that do not meet the standards
\end{itemize}

For outputs that fail to meet these criteria, specific issues are identified, and the model is instructed to regenerate outputs that adhere to the required standards based on the evaluation results.

\subsection{Secondary review by an independent model} 
Following the initial validation, a second large language model, Gemini-1.5-pro, is employed to independently review and verify the accuracy and consistency of the reasoning paths. Additionally, GPT4o-mini is tasked with reconstructing any unqualified cases based on feedback from Gemini-1.5-pro.

This rigorous quality assurance process not only ensures the integrity of the data but also lays a strong foundation of high-quality training data, enhancing interpretability in downstream tasks.

\subsection{Final review by MEME tool and human}
To further corroborate our findings, we utilized the MEME tool (Multiple Expectation Maximizations for Motif Elicitation) to identify motifs in DNA sequences. The motifs discovered by MEME showed strong consistency with the reasoning data, providing additional validation. We also invited biology PhD students and professors to conduct human evaluations of the reasoning data. Their feedback confirmed the high quality and reliability of the generated data. An example of our data quality control process is shown in Table~\ref{quality control process example}.

\label{Appendix:B}

\begin{figure*}[t]
\begin{center}
\includegraphics[width=\textwidth]{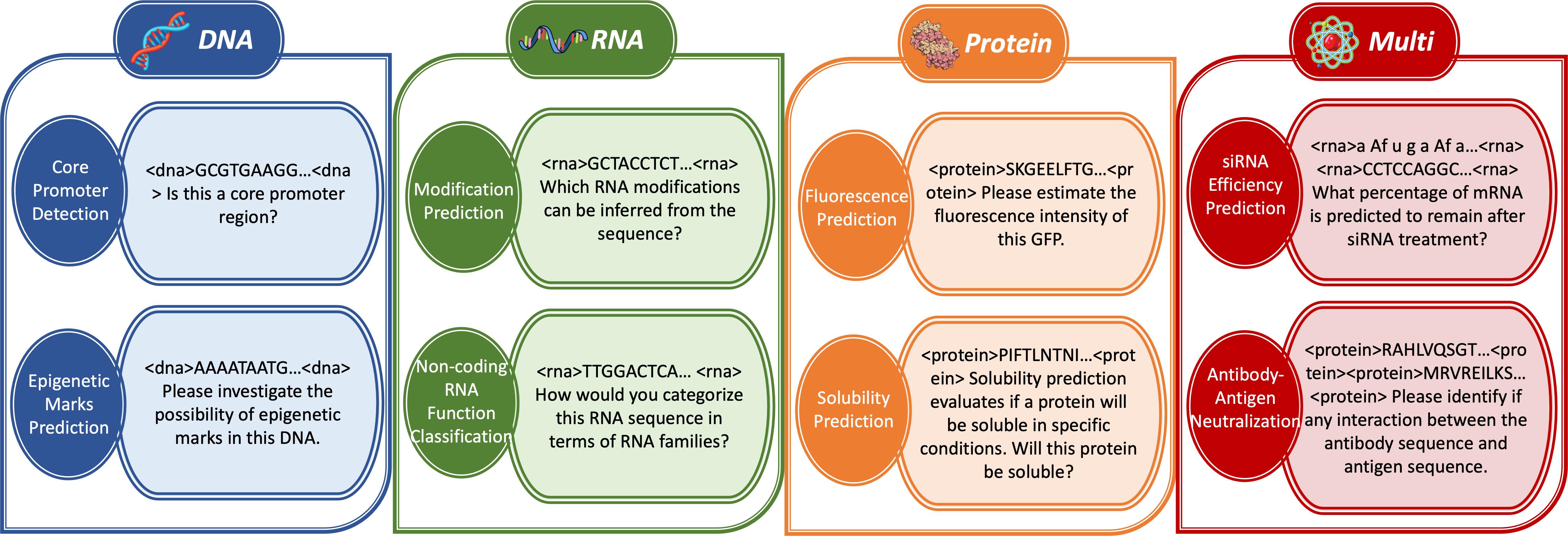}
\end{center}
\vspace{-3mm}
\caption{Examples of instruction prompts constructed for each omics type.}
\label{fig: instruction prompts}
\vspace{-3mm}
\end{figure*}

\begin{table*}[ht]
\caption{Examples of question and answer template pairs in stage 2 training data.}
\centering
\resizebox{\textwidth}{!}{
\begin{tabular}{>{\raggedright}p{0.25\textwidth} p{0.4\textwidth} p{0.35\textwidth}}
\toprule
\textbf{Task} & \textbf{Question template} & \textbf{Answer template} \\
\midrule
Epigenetic Marks Prediction & \texttt{<dna>\{DNA\}</dna>} Are there any characteristic epigenetic marks in this DNA? & After careful EMP analysis, there is conclusive evidence of epigenetic marks in the given DNA sequence. (Positive case) \\
Core Promoter Detection & \texttt{<dna>\{DNA\}</dna>}: Evaluate this sequence for potential promoter regions. & No, a promoter region is not present in the given genomic fragment. (Negative case) \\
Enhancer Activity Prediction & \texttt{<dna>\{DNA\}</dna>} Enhancer activity in this sequence - what's the deal? & The enhancer activity prediction yields: HK - \{hk\_enrichment\}, Dev - \{dev\_enrichment\} \\
CRISPR On Target Prediction & \texttt{<rna>\{RNA\}<rna>} What is the predicted on-target activity of the sequence? & The sequence has an on-target efficiency score of \{label\}.\\
Programmable RNA Switches & \texttt{<rna>\{RNA\}<rna>} What ON/OFF ratio can be expected from the sequence? & The ON state for this sequence is \{label\_ON\}, the OFF state is \{label\_OFF\}, and the ON/OFF ratio is \{label\_ON\_OFF\}. \\

Modification Prediction & \texttt{<rna>\{RNA\}<rna>} [RNA modification classification] Which RNA modifications are inferred from the sequence? & The RNA modifications for the sequence are predicted as \{label\}. \\
Fluorescence Prediction & \texttt{<protein>\{protein\}<protein>} How effective is the predicted fluorescence for this GFP sequence? & The GFP's fluorescence is calculated to be \{label\}. \\
Enzyme Commission Number Prediction & \texttt{<protein>\{protein\}<protein>} What is the specific activity of this protein as an enzyme? & Specific enzyme activities are defined by EC number \{label\}. \\
Solubility Prediction & \texttt{<protein>\{protein\}<protein>} Can solubility be confirmed for this protein? & This protein shows positive solubility. (Positive case) \\
Enhancer-Promoter Interaction Prediction & \texttt{<dna>\{DNA1\}<dna>} promoter + \texttt{<dna>\{DNA2\}<dna>} enhancer: connected? & We find no support for promoter-enhancer communication in this genetic material. (Negative case) \\
Antibody-Antigen Neutralization & Is there any molecular recognition between this Ab-Ag pair: \texttt{<protein>\{AB\}<protein>>}-\texttt{<protein>\{AG\}<protein>>}? & The computational predictions do not support any significant Ab-Ag interaction. (Negative case) \\

siRNA Efficiency Prediction & \texttt{<rna>\{SiRNA\}<rna>} \texttt{<dna>\{targetDNA\}<dna>} How does the target RNA perform in terms of siRNA-induced mRNA reduction? & The siRNA treatment results in an mRNA remaining percentage of \{label\}.\\
\bottomrule
\end{tabular}
}
\label{table:templates}
\end{table*}

\begin{table*}[ht]
\caption{An example of question and answer pairs in stage 3 training data for siRNA Efficiency Prediction task.}
\centering
\resizebox{\textwidth}{!}{
\begin{tabular}{p{0.2\textwidth} p{0.75\textwidth}}
\hline
\multicolumn{2}{c}{\textbf{Task: siRNA Efficiency Prediction}} \\ \hline
\textbf{Question} & \texttt{<rna>\{SiRNA\}<rna>} \texttt{<dna>\{targetDNA\}<dna>} How much mRNA remains after siRNA treatment of the target RNA? \\ 
\\
\textbf{Answer} & To assess the amount of mRNA remaining after siRNA treatment, we first need to understand the mechanisms involved in RNA interference (RNAi) and its impact on mRNA levels. \newline \newline
siRNA (small interfering RNA) is a class of double-stranded RNA that can lead to the degradation of complementary mRNA molecules, effectively silencing their expression. When siRNA binds to target mRNA, it recruits the RNA-induced silencing complex (RISC), which then cleaves the mRNA, reducing its stability and leading to its degradation. \newline \newline
To determine how much mRNA remains after treatment, one typically considers: \newline
1. The initial concentration of mRNA. \newline
2. The efficacy of siRNA in degrading the target mRNA. \newline
3. The kinetic parameters related to both the mRNA decay rate and the siRNA efficacy. \newline \newline
For instance, if initial studies or assays show that treatment with specific siRNA reduces target mRNA levels by approximately 80\% under standard experimental conditions, then we can assume that 20\% of the original mRNA remains. However, the problem states that the final calculated remaining mRNA is estimated to be 80.6\% of the original mRNA. \newline \newline
This indicates that the siRNA treatment was not completely effective, or there might be factors such as incomplete targeting or secondary structures in the mRNA that protected it from degradation. \newline \newline
In summary, after siRNA treatment of the target RNA, the remaining amount of mRNA is therefore 80.6\%, indicating a less than complete silencing effect. This percentage reflects the balance of mRNA decay, the effectiveness of the siRNA treatment, and the biological context in which the treatment was performed. \\ \hline
\end{tabular}
}
\label{table:stage3_siRNA}
\end{table*}

\begin{table*}[ht]
\caption{An example of question and answer pairs in stage 3 training data for Noncoding RNA Family Prediction task.}
\centering
\resizebox{\textwidth}{!}{
\begin{tabular}{p{0.2\textwidth} p{0.75\textwidth}}
\hline
\multicolumn{2}{c}{\textbf{Task: Noncoding RNA Family Prediction}} \\ \hline
\textbf{Question} & \texttt{<rna>}\{RNA\}\texttt{<rna>} How does this RNA sequence contribute to gene regulation? \\ 
\\
\textbf{Answer} & This RNA sequence represents a component that likely plays a crucial role in gene regulation, particularly through its classification as a "leader" RNA. In the context of molecular biology, leader RNAs are typically found at the 5' end of mRNA sequences and can influence gene expression by engaging in regulatory interactions with ribosomes or other molecular partners. \newline \newline
Leader sequences can have multiple functions, including: \newline
1. **Regulatory Functions:** They may harbor sequences that can bind small regulatory molecules or proteins, thus modulating the transcription or translation of the downstream coding region. \newline
2. **Secondary Structures:** The folding patterns in leader sequences can form intricate secondary structures that contribute to transcriptional attenuation or the formation of a riboswitch. These structures can determine whether the RNA remains in a transcriptionally active state or folds into a configuration that halts transcription or translation. \newline
3. **Interaction with Ribosomes:** The leader region often acts as a signal for ribosomes to initiate translation. For mRNAs, the presence of specific motifs within this region can influence how efficiently translation occurs, impacting overall protein synthesis levels and thus the expression of the associated gene. \newline \newline
In summary, the RNA sequence presented can be classified as a leader, contributing to gene regulation through its interactions and structural characteristics. This underscores the importance of untranslated regions (UTRs) in controlling gene expression. Understanding such RNA sequences and their functions provides valuable insights into the mechanisms of gene regulation, highlighting how sequences beyond coding regions play vital roles in cellular processes. \\ \hline
\end{tabular}
}
\label{table:stage3_noncodingRNA}
\end{table*}

\begin{table*}[ht]
\caption{An example of question and answer pairs in stage 3 training data for Core Promoter Detection.}
\centering
\resizebox{\textwidth}{!}{
\begin{tabular}{p{0.2\textwidth} p{0.75\textwidth}}
\hline
\multicolumn{2}{c}{\textbf{Task: Core Promoter Detection}} \\ \hline
\textbf{Question} & \texttt{<dna>}\{DNA\}\texttt{<dna>} Identify potential core promoter region in this sequence. \\ 
\\
\textbf{Answer} & The provided DNA sequence shows characteristics that warrant careful examination for core promoter elements. Core promoters are essential in initiating transcription and are typically located upstream of the transcription start site. They often contain specific motifs, such as the TATA box, initiator sequences (Inr), and other regulatory elements that facilitate the binding of transcription factors and RNA polymerase II. \newline \newline
In the sequence provided, critical components of core promoters might include: \newline
1. **Initiator sequences (Inr):** These sequences often span the transcription start site and can exhibit variability but typically include consensus elements that help delineate the start of transcription. \newline
2. **TATA box:** A common feature among many eukaryotic promoters, which is usually located about 25-30 bases upstream from the transcription start point. \newline
3. **GC-rich regions:** These regions can also play a role in promoter activity by providing binding sites for transcription factors. \newline \newline
Upon examining the sequence: \newline
- There are no clear TATA boxes or Inr motifs that align with typical core promoter signatures. \newline
- The sequence appears GC-rich, notably towards the middle, but does not show significant promoter elements consistent with those typically required for core promoter identification. \newline \newline
Given these observations, we can conclude that this sequence does not contain recognizable features indicative of a core promoter region. Therefore, the response to whether a potential core promoter region is present in this sequence is negative. \\ \hline
\end{tabular}
}
\label{table:stage3_corePromoter}
\end{table*}

\begin{table*}[ht]
\caption{An example of quality control process.}
\label{quality control process example}
\centering
\resizebox{\textwidth}{!}{
\begin{tabular}{p{0.2\textwidth} p{0.8\textwidth}}
\hline
\multicolumn{2}{c}{\textbf{Task: Transcription Binding Sites Detection}} \\ \hline
\textbf{Question} &\texttt{<dna>}AGTTGC......AGGCCC\texttt{<dna>}: Can you check if this DNA fragment contains sequences associated with transcription initiation? \\ 
\\
\textbf{Answer} & The provided DNA fragment showcases a sequence that can be analyzed to determine if it contains motifs associated with transcription initiation. Transcription initiation in eukaryotic organisms often involves the presence of specific cis-regulatory elements, such as promoter regions, which may include TATA boxes, GC-rich regions, or other binding sites for transcription factors.
\newline
In the analysis of the given sequence, we should look for characteristic motifs common in eukaryotic promoters:
\newline
TATA Box: Usually located 25-30 base pairs upstream of the transcription start site, this sequence often resembles the consensus sequence TATAAA.
\newline
GC Box: These sequences are rich in GC content and provide binding sites for transcription factors (e.g., Sp1). They typically appear around -100 to -200 base pairs upstream.
\newline
CCAAT Box: Another common element involved in the binding of transcription factors, found further upstream, approximately -75 bases from the start site.
\newline
Upon examining the sequence you provided, several regions exhibit characteristics of transcriptional regulation:
\newline
The sequence includes stretches of CG-rich content that might suggest the presence of a GC box. Additionally, there is a notable absence of any clear TATA box or canonical elements, which indicates that while the segment may hold potential as a regulatory region, it may not conform to traditional promoter motifs rigorously. In summary, based on the detection of indirect evidence of transcription initiation sites, particularly the presence of GC-rich sequences, we can conclude that this DNA fragment likely contains sequences associated with transcription initiation. Therefore, the answer to your question is affirmative; it is positive that this DNA fragment likely includes sequences relevant to the initiation of transcription. Further studies, such as promoter assays or bioinformatics tools, could further validate this observation. \\
\\
\textbf{AI Evaluation} &
GPT4o-mini analysis: The answer provides a detailed analysis of the DNA sequence with specific references to hallmark motifs and features associated with promoters. It incorporates relevant biological knowledge and presents a clear conclusion regarding the presence of promoter-like characteristics.
\newline \newline
Gemini-1.5-pro evaluation: The answer provides a comprehensive analysis of the DNA sequence for promoter-like features. It correctly identifies key characteristics such as GC-rich regions and potential transcription factor binding sites. The reasoning is clearly explained, starting with a general description of promoters and then moving to specific features found within the provided sequence. The conclusion is clearly stated and supported by the analysis. While it doesn't provide specific locations of potential promoter elements or quantify the GC content, the overall analysis is sufficient for a general assessment of promoter-like characteristics.\\
\\
\textbf{Human Evaluation} &
For promoters, our human evaluation involved examining the DNA sequences, where we observed a typical TATA box motif at positions -31 to -35, aligning well with the reasoning for TATA-type promoters. Additionally, we utilized the MEME tool (Multiple Expectation Maximizations for Motif Elicitation), which identifies motifs in related DNA sequences. The motifs discovered by MEME also showed strong consistency with the reasoning, further validating our findings.
\\ \hline
\end{tabular}
}
\end{table*}

\end{document}